\documentclass[useAMS,usegraphicx,usenatbib]{mn2e}
\usepackage[a4paper,centering, totalwidth=520pt, totalheight=700pt]{geometry}
\usepackage[position=top,labelformat=empty]{subfig}
\usepackage{graphicx}
\usepackage{aas_macros}     
\usepackage{amsmath}
\usepackage{amssymb}
\usepackage{fixltx2e}[1999/12/01]
\usepackage{tikz}
\usepackage{txfonts}

\newcommand{\Msun}{\ensuremath{M_{\odot}}}
\newcommand{\Mpc}{ \ensuremath{h^{-1} {\rm Mpc}} }

\newcommand{\kpc}{ \ensuremath{ h^{-1} {\rm kpc}} }

\newcommand{\mps}{ m_{\rm p}} 
 
\newcommand{\Rvir}{ R_{\rm vir}} 
\newcommand{\Mvir}{\rm M_{vir}} 
 
\newcommand{\dd}{{\rm d}}

\DeclareMathOperator\erfc{erfc}
\def\simlt{\lower.5ex\hbox{$\; \buildrel < \over \sim \;$}}
\def\simgt{\lower.5ex\hbox{$\; \buildrel > \over \sim \;$}}

\title[Neutralino haloes]{Earth-mass haloes and the emergence of NFW density profiles}

\begin{document}
\setlength{\topmargin}{-1.cm}

\author[Angulo et al.]{
\parbox[h]{\textwidth}
{Raul E. Angulo$^{1} \thanks{rangulo@cefca.es}$,
Oliver Hahn$^{2}$,
Aaron D. Ludlow$^{3}$ \&
Silvia Bonoli$^{1}$} 
\\
\\
$^1$ Centro de Estudios de F\'isica del Cosmos de Arag\'on (CEFCA), Plaza San Juan 1, Planta-2, 44001, Spain. \\
$^2$ Laboratoire Lagrange, Universit\'e C\^ote d'Azur, Observatoire de la C\^ote d'Azur, CNRS, \\
Blvd de l'Observatoire, CS 34229, 06304 Nice cedex 4, France.\\
$^3$ Institute for Computational Cosmology, Dept. of Physics, Univ. of Durham, South Road, Durham DH1 3LE, UK
}
\date{submitted to MNRAS}
\maketitle

\pagerange{\pageref{firstpage}--\pageref{lastpage}} \pubyear{2016}
\label{firstpage}

\begin{abstract} 
We simulate neutralino dark matter ($\chi$DM) haloes from their initial collapse, at
$\sim$ earth mass, up to a few percent solar. Our results confirm that the density
profiles of the first haloes are described by a $\sim r^{-1.5}$ power-law. As
haloes grow in mass, their density profiles evolve significantly. In the
central regions, they become shallower and reach on average $\sim r^{-1}$, the
asymptotic form of an NFW profile. Using non-cosmological controlled
simulations, we observe that temporal variations in the gravitational potential
caused by major mergers lead to a shallowing of the inner profile. This
transformation is more significant for shallower initial profiles and for a
higher number of merging systems. Depending on the merger details, the
resulting profiles can be shallower or steeper than NFW in their inner regions.
Interestingly, mergers have a much weaker effect when the profile is given by a
broken power-law with an inner slope of $-1$ (such as NFW or Hernquist
profiles). This offers an explanation for the emergence of NFW-like profiles:
after their initial collapse, $r^{-1.5}$ $\chi$DM haloes suffer copious major
mergers, which progressively shallows the profile. Once an NFW-like profile is
established, subsequent merging do not change the profile anymore.  This
suggests that halo profiles are not universal but rather a combination of (1)
the physics of the formation of the microhaloes and (2) their early merger
history -- both set by the properties of the dark matter particle -- as well as
(3) the resilience of NFW-like profiles to perturbations.  \end{abstract}

\section{Introduction} 

Soon after the first $N$-body simulations reached sufficient mass and force
resolution, they revealed that the spherically-averaged density profiles of
cold dark matter (CDM) haloes have a central density cusp with a sharp decline
towards their outskirts \citep[e.g][]{Frenk1988,Dubinski1991,Gelb1994}. Overall,
the density profiles appeared to be well described by a very simple functional
form,

\begin{equation}
\rho(x) \propto \frac{1}{x} \frac{1}{(1+x)^2},
\label{eq:NFW}
\end{equation}

\noindent \citep[][hereafter the NFW profile]{Navarro1996,navarro1997}
regardless of halo mass, cosmological parameters or the details of the
fluctuation power spectrum
\citep[e.g.][]{Cole1996,Huss1999,Bullock2001,Wang2007,Lovell2012}. Even today,
after decades of advances in computational power and techniques -- with
improvements of over a factor of $10^{5}$ in particle number and $10^2$ in
force resolution \citep{Diemand2007, Springel2008, Gao2012} -- the universality
and the asymptotic slope of the density profile of $\sim-1$ still holds with
only minor corrections \citep[e.g.][]{Navarro2010}. This is arguably one of the
most important results in computational cosmology to date, yet it is puzzling
when contrasted with the expectations of early analytic treatments of the cold
collapse of primordial fluctuations
\citep{GunnGott1972,Fillmore1984,Bertschinger1985}.  In particular, these
models predict simple power-law density profiles whose slope depends
sensitively on that of the initial collapsing patch.

Although the origin of the NFW shape and its universality are still under debate
there has been no lack of attempts to explain it. For example,
\cite{Ludlow2013} argued that, when expressed in appropriate units, the NFW
profile is indistinguishable from that of the scale-free shape of CDM halo mass
accretion histories. In this interpretation, the NFW profile simply reflects
the typical background density-dependence of mass accretion onto growing dark
matter haloes.  Other authors apply principles of maximum entropy or adiabatic
invariance to show how profiles similar to eq.~(\ref{eq:NFW}) may result from
strong mixing associated with mergers and accretion occurring during halo
assembly \citep[e.g.][]{TN2001,Dalal2010,Pontzen2013,Juan2014}, or highlight
the importance of angular momentum in driving profiles toward the NFW shape
\citep[e.g.][]{Lentz2016}. Others speculate whether the graininess of the
$N$-body method might spuriously drive the profiles to the NFW-shape
\citep[e.g.][]{Baushev2015}.

Simulations that resolve the free-streaming scale of the dark matter (DM) particle may shed light
on the physical origin of DM halo density profiles. The free-streaming scale is
the only relevant scale in DM cosmologies: it initiates a well-defined
and finite hierarchy of structure and implies that the first haloes can be
numerically resolved given adequate resolution. This is quite different from
``standard'' CDM simulations (i.e. in the perfectly cold limit), where DM 
traces perturbations to arbitrarily small scales such that there 
are always density perturbations in the DM at the resolution limit. 
In this case, the first haloes are resolved with just a few particles and thus can
be heavily influenced by numerical noise which is arguably at least of comparable
magnitude as the physical perturbations on mean inter-particle scales.

For the lightest neutralino in supersymmetric extensions to the standard model of
particle physics, which should have a mass close to $100\,{\rm GeV}$, the free streaming scale corresponds to
about an earth mass \citep{Hofmann2001,Green2004,Diemand2005b}.  While it is computationally impossible given today's
resources to simulate the full hierarchy of DM haloes from micro-halo
to galactic scales, it {\em is} possible to simulate the formation of the first
haloes by focusing on high redshifts and small volumes
\citep[e.g.][]{Diemand2005b}. Such simulations have had a rather unexpected
outcome: the density profiles of micro-haloes appear to deviate substantially
from the NFW shape. Instead, they are better described by a single, steep
power-law, $\rho \propto r^{-1.5}$ \citep{Diemand2005b,
Ishiyama2010,Anderhalden2013}. This appears to be a stable configuration and
not simply a transient product of a rapid mass accretion
\citep[see][]{Ishiyama2014}. In addition, \cite{Ishiyama2014} found that the
average profiles of haloes approach the NFW form as their masses grow well
above the free-streaming scale.  Conversely, for a perfectly cold CDM spectrum 
without a cut-off, haloes {\em always} exhibit an NFW profile, regardless of their 
mass.

Clearly, the discussion above raises several important questions: What is the
physics that determines the density profiles of the first generation of dark
matter haloes?  Why do single power-laws describe well some haloes, while
others are better described by gently curving profiles, such as the NFW form?  What is
the mechanism that transforms the initial power-law profiles of neutralino
haloes to NFW-like profiles at later times? What is the role of their assembly
history in establishing the initial profile shape and in its subsequent
evolution? Are (at least) some of these results driven by numerical noise and
convergence problems of the $N$-body method? All these questions need to
be answered to obtain a complete picture of the physics behind the equilibrium
(or, at least, stable) state of collapsed objects in the Universe.

In this paper, we directly address some of these issues using zoom simulations of six
neutralino DM haloes, which we follow from their first collapse at the
free-streaming scale until they have reached $\sim0.02\Msun$. We show that
haloes initially have power-law profiles, $\rho \propto r^{-1.5}$, a form which
is preserved during periods of smooth accretion. However, we find that our 
haloes undergo a transformation in which their density profiles move
toward a broken power-law form. We investigate a possible origin of this
transformation using a suite of idealised, non-cosmological merger simulations.
We find clear evidence that a flattening of the central slope is a natural
outcome of a fluctuating potential produced by mergers. However, these
simulations also show that the final profile depends strongly on the number
as well as on the initial profiles of the merging systems. In particular, NFW-like 
profiles appear
significantly more stable to such strong perturbations than simple power-laws.
Although there are still open questions, these results offer useful insights 
into the physical processes relevant for the emergence of universal density
profiles.

The outline of this article is as follows. First, in \S2, we describe our
cosmological simulations of neutralino-DM haloes and the main aspects
of our analysis, followed by a discussion of the density profiles of these
objects in \S2.3. In the second part, we focus on non-cosmological
simulations. In \S3 we detail the methods employed for generating different
equal-mass merger configurations and then discuss the simulation results,
focusing mainly on the transformation of density profiles due to temporal
oscillations of the central potentials. Finally, in \S4 we discuss and
interpret our results in a broader context and provide a concluding 
discussion.

\section{Numerical Simulations}

We use cosmological simulations to study the formation histories and internal
structure of the first haloes to collapse in a universe dominated by
neutralino DM.  All of our runs adopt the following cosmological
parameters: $\Omega_m=0.3$, $\Omega_{\Lambda}=0.7$, $h = 0.7$, $n_s = 1$, and
$\sigma_8 = 0.87$.  Here, $\Omega_i$ refers to the current energy density of
component $i$ in units of the critical density, $\rho_{\rm crit}$; $h$ is the
Hubble parameter, expressed in units of $100 \, {\rm km}\, s^{-1}{\rm Mpc}$;
$n_s$ is the spectral index of primordial density perturbations, and $\sigma_8$
is the rms density fluctuation measured in 8 Mpc spheres, linearly
extrapolated to $z=0$.  We adopt a transfer function consistent with a standard
neutralino particle of mass $m_{\chi} = 100$~GeV, with a decoupling temperature
of ${\rm T}=33$~MeV \citep[see, e.g.,][]{Green2004}, which we have implemented
in the publicly-available {\tt MUSIC} code \citep{Hahn2011a}. The resulting
linear theory power spectrum is shown in Fig.~\ref{fig:pk} using a solid line. 

In such a cosmology, the free-streaming of neutralino particles in the early Universe 
suppresses the growth of density
perturbations below a characteristic scale of $\sim 0.7$ pc. This implies that the
very first objects to collapse (which are also the most abundant) do so at $z >60$ 
and have masses of order $10^{-6}\,{\rm M}_{\odot}$
\citep{Green2004,Diemand2005b,Angulo2010}.  Due to the enormous dynamic range
required, it has not yet been possible to carry out self-consistent $N$-body
simulations that follow the full hierarchical growth of DM haloes from the
free-streaming mass up to galactic scales. Furthermore, the need to also properly
model large-scale modes in a DM model has hampered efforts to simulate
neutralino haloes of the lowest mass from their formation epoch to the present
day. For that reason, their internal structure and spatial distribution within
the haloes of galaxies like the Milky Way has not yet been properly
characterized.

Here, we will perform a simulation of a relatively large region at high
redshift and use this run to identify and excise objects for re-simulation at
much higher resolution.  Doing so will allow us to resolve the initial collapse
at the free-streaming scale, and to follow in detail the subsequent evolution
of these haloes as they grow in mass by several orders of magnitude. In the
following subsections we summarise the relevant details of these simulations.

\subsection{Numerical setup}

\subsubsection{The parent simulation}

\begin{figure}
\includegraphics[width=8.5cm]{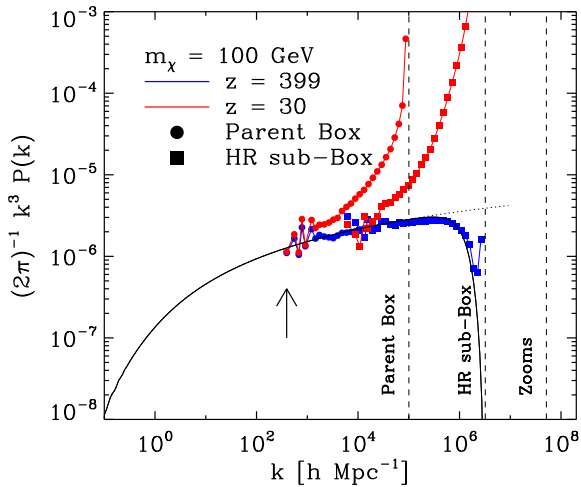}
\caption{The linear power spectrum of mass fluctuations assuming dark matter to
be composed of neutralinos of mass $m_{\chi}=100$ GeV\label{fig:pk}. The solid
line corresponds to the linear theory power spectrum employed to set up our
simulations; the dotted line shows, for comparison, the power
spectrum of a CDM model without a small-scale cut-off. Red and blue symbols indicate the
power spectra measured in our parent simulation at $z=30$ and $z = 399$,
respectively, both linearly scaled to $z=0$. In both cases, circles indicate
the spectra of the whole $16\,\kpc$ parent box, whereas squares that of the central 
$1\,\kpc$ box followed at higher resolution. The vertical arrow shows the
fundamental mode of our simulations, and the vertical dashed lines mark
the particle Nyquist frequency for the whole parent box, high-resolution 
sub-box, and for the zoom simulations. \label{fig:pk}}
\end{figure}

Initial conditions for our parent simulation use $512^3$ particles of mass
$7.4\times10^{-7}\Msun$ to sample the linear density field within a cubic
region of side-length $L=1.4$ kpc, which is itself embedded into a larger,
lower-resolution, periodic box of $L=22.85$ kpc. Particle positions and
velocities are generated using 2nd-order Lagrangian Perturbation Theory at a
starting redshift of $z=399$, which is sufficiently high to ensure that
perturbations on the smallest resolved scales are well within the linear
regime. Gravitational evolution was simulated using a Tree+PM algorithm
implemented in the {\tt P-Gadget3} code \citep{Springel2008}, down to a 
final redshift of $z=30$ and using a softening length parameter $\epsilon=0.1$
pc.

The power spectrum measured at the starting redshift is shown in
Fig.~\ref{fig:pk} using blue symbols, which agrees well with the linear theory
prediction. Note the characteristic suppression of fluctuations on scales
$k\simgt 10^6\,\Mpc^{-1}$. On these scales the logarithmic slope of the power
spectrum is close to $-3$, and fluctuations on a broad range of spatial scales
are expected to collapse simultaneously (this is due to a similar amplitude of
the mass variance on those scales). Halo growth is therefore initially very
rapid and, shortly after the first objects collapse, the non-linear scale
approaches the size of the box. As a result, large boxes are required to
properly simulate the evolution of the first neutralino DM haloes. In our case,
the parent box-length is $\sim 3000$ times larger than the virial radius of the least
massive target halo. As we can see from a comparison of the initial (blue
symbols) and final (red symbols) power spectra, only the four lowest modes are
still well described by linear theory. Thus, by the final output, the
non-linear scale is comparable to (but still smaller than) the box-size. We
note, however, that missing large-scale modes plausibly affect the time of
collapse of our haloes to some degree; boxes of several cubic Mpcs or larger
are required to achieve full convergence \citep{Ishiyama2010}. Nevertheless,
the formation and evolution of haloes will be qualitatively correct: the
gravitational collapse of a cold fluid with a given initial density spectrum,
followed by multiple mergers and smooth mass accretion. Note also the offset between the spectra of
the parent box (circles) and that of the high-resolution sub-box (squares), which
is caused by the latter being a relatively underdense region.

\begin{figure*} 
\begin{center}
\subfloat{\includegraphics[width=0.49\textwidth]{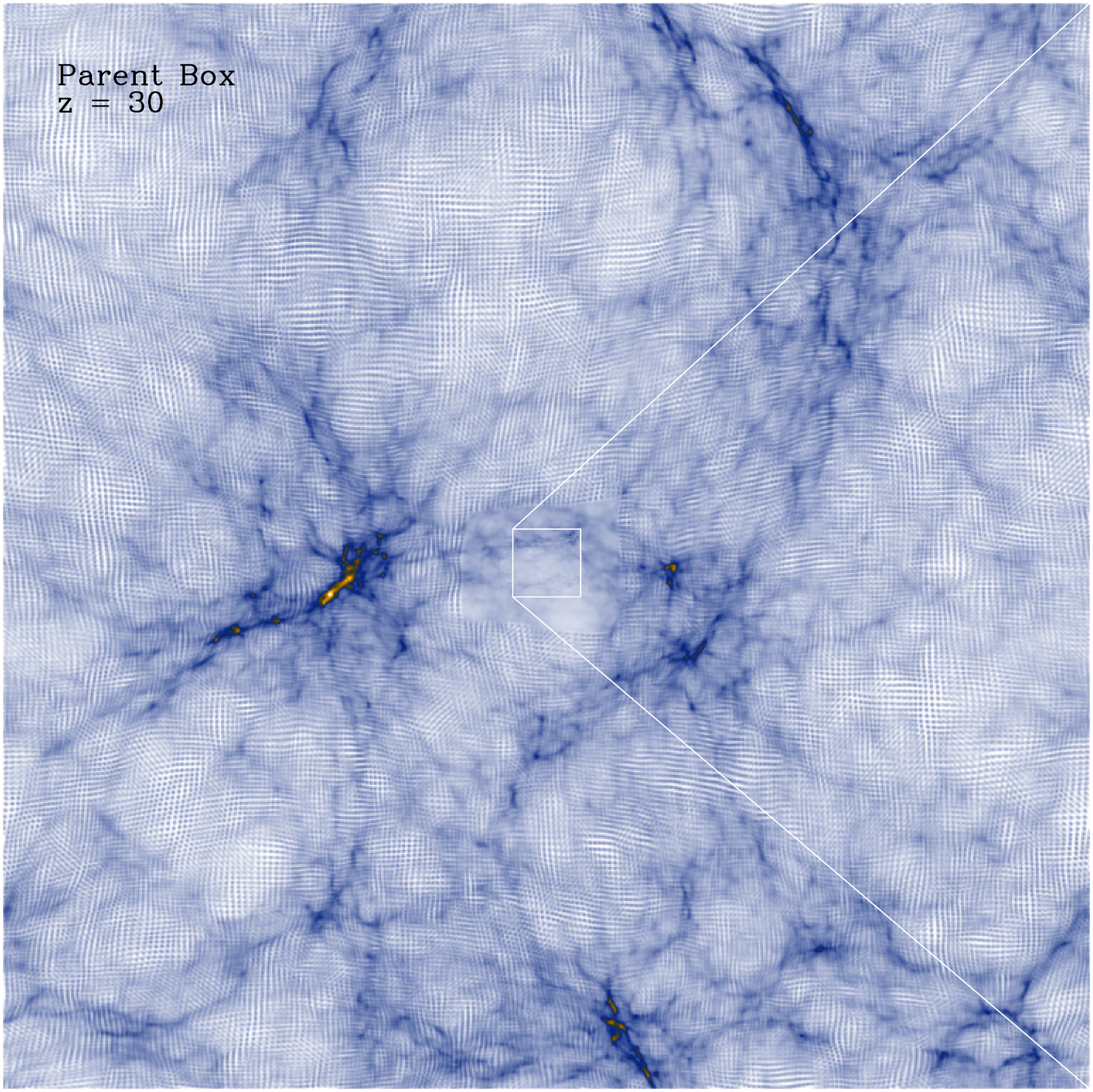} }
\subfloat{\includegraphics[width=0.49\textwidth]{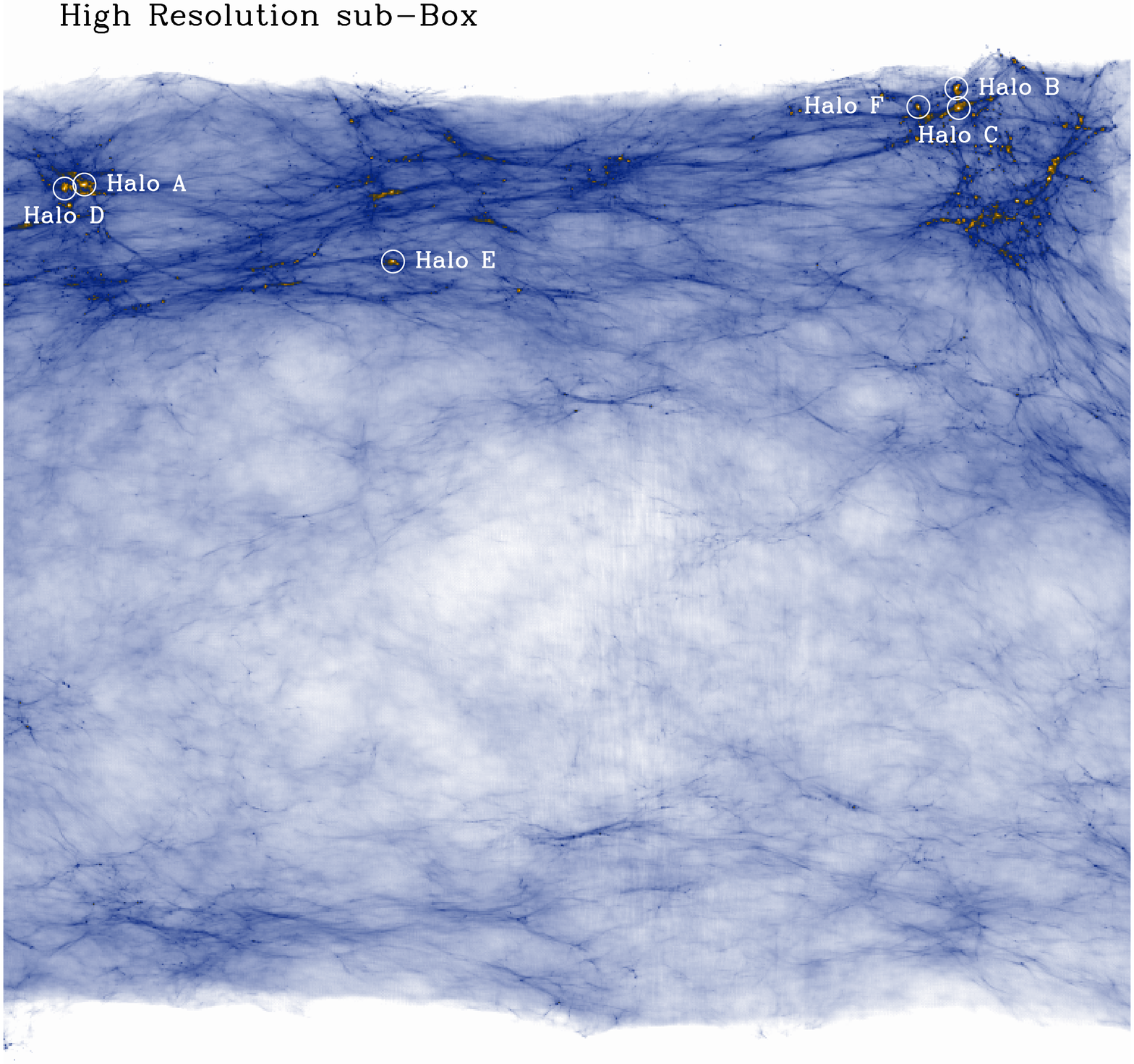} }
\caption{The projected density of dark matter in our parent simulation at
$z=30$. The image displays a $1400$ pc thick slab of a the whole parent
box ($22\times22$ kpc; left panel), and of the high-resolution sub-region 
($1.4\times1.4$ pc; right panel), 
oriented such that they contains all 6 haloes in our
re-simulation suite (highlighted by white circles). Note the overall sparcity of
collapsed structure resulting from the truncation in the initial power spectrum
caused by the free streaming of neutralinos. \label{fig:full}}
\end{center}
\end{figure*}

The above considerations can be visually appreciated in Fig.~\ref{fig:full},
which shows the density field at the final output of our simulation. Clearly
visible are very extended regions where fluctuations have not yet fully
collapsed. These give rise to an incipient network of filaments and only a
handful of highly-clustered haloes (corresponding to the rarest peaks in the
simulation volume) can be clearly identified (indeed, at this redshift only
$0.7\%$ of the mass is associated with {\em any} halo). The surroundings of
this region are also over-dense, which emphasises the fact that multiple scales
reach the threshold for collapse at similar times. 

\subsubsection{The resimulation suite}

\begin{figure*} 
\subfloat{\includegraphics[width=0.33\textwidth]{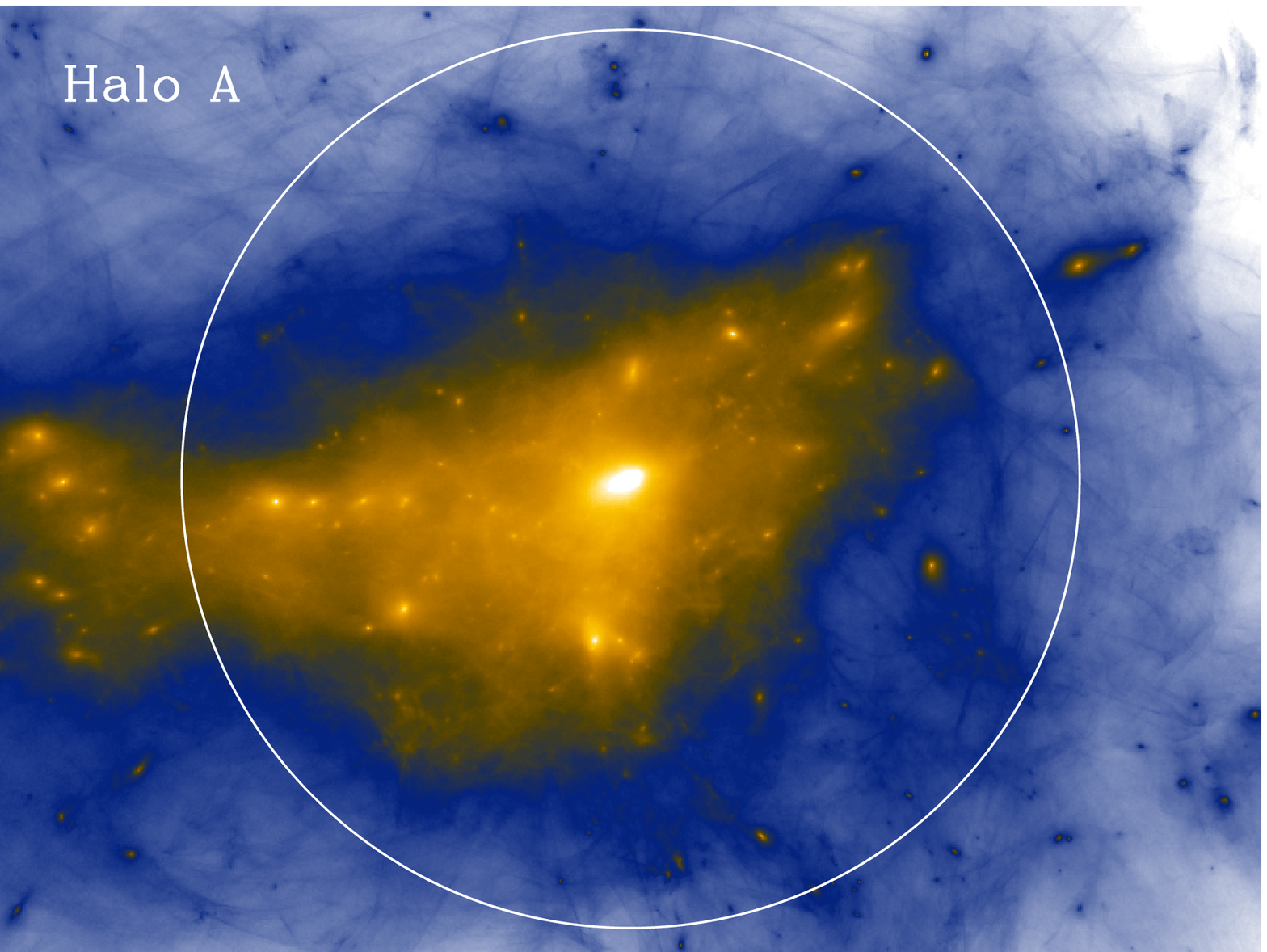} }
\subfloat{\includegraphics[width=0.33\textwidth]{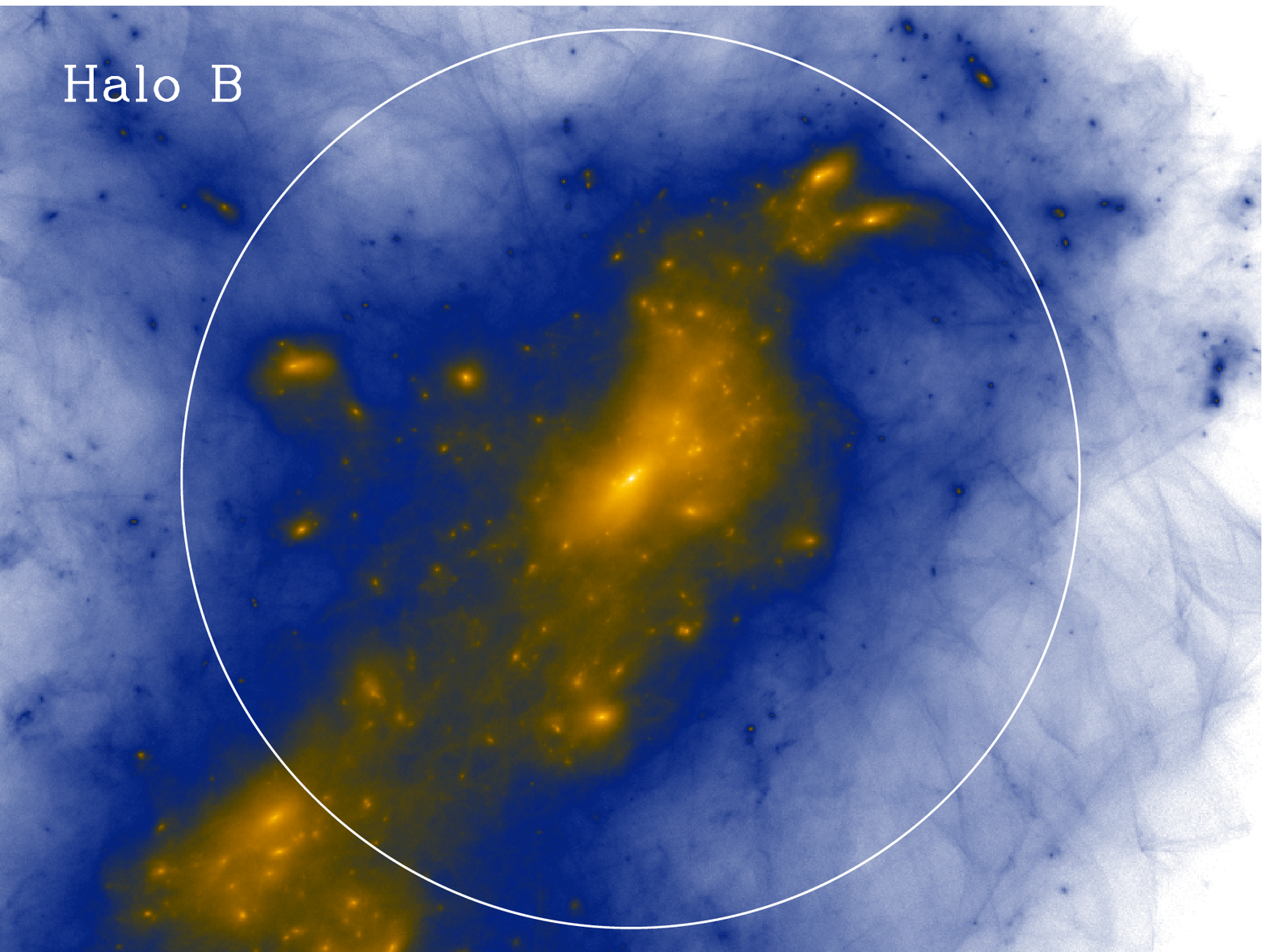} }
\subfloat{\includegraphics[width=0.33\textwidth]{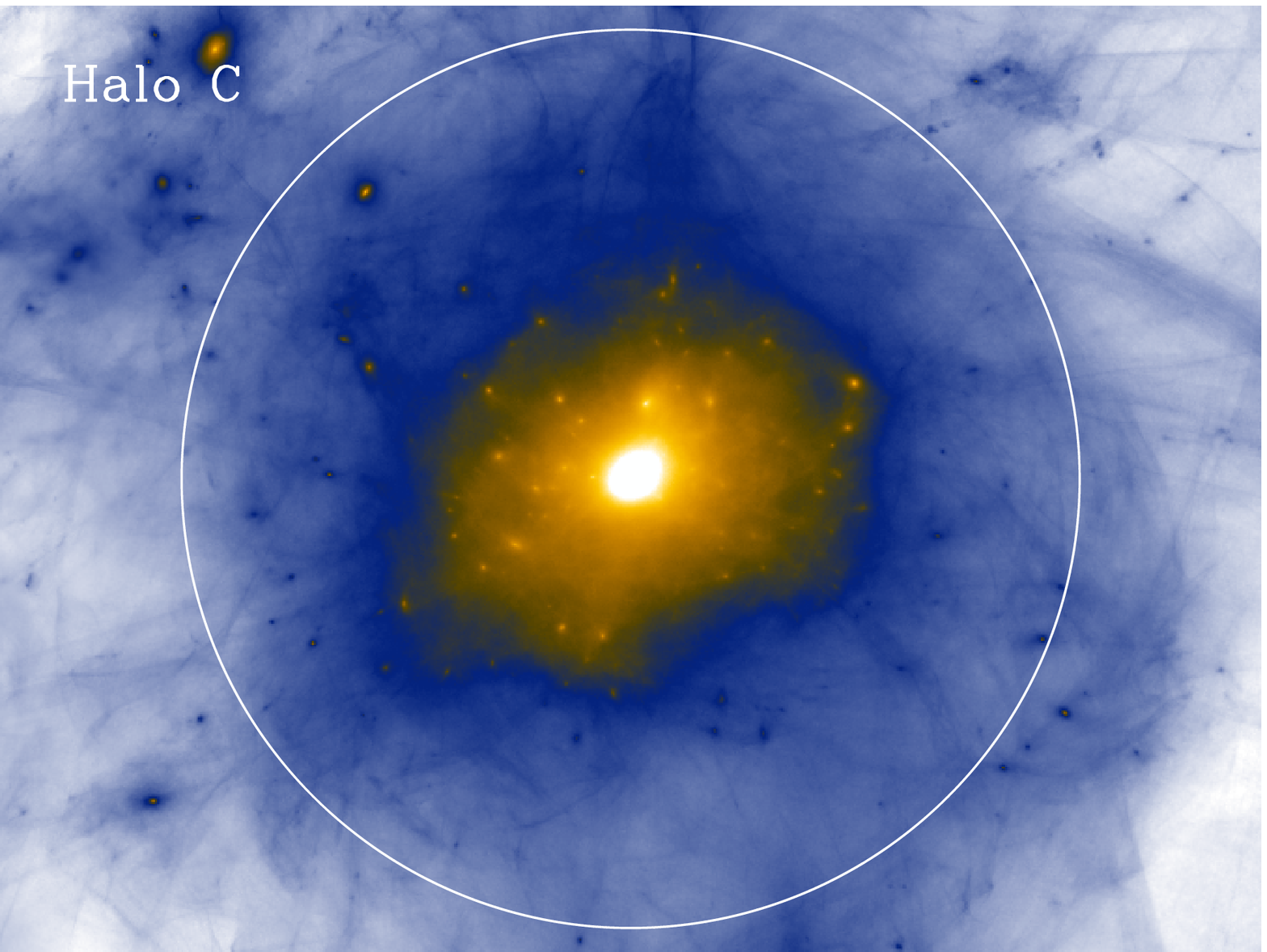} } \\[-2ex]
\subfloat{\includegraphics[width=0.33\textwidth]{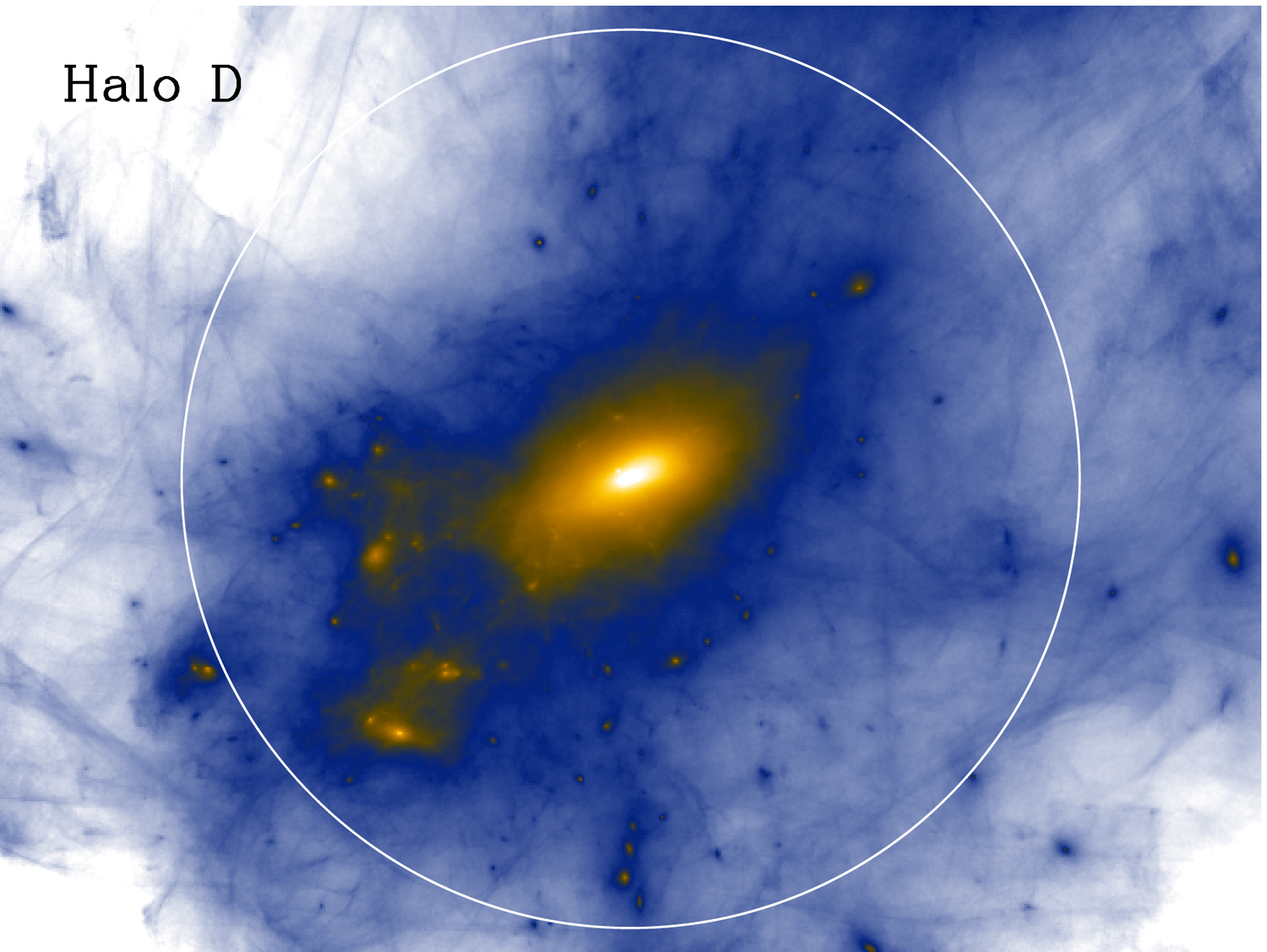} } 
\subfloat{\includegraphics[width=0.33\textwidth]{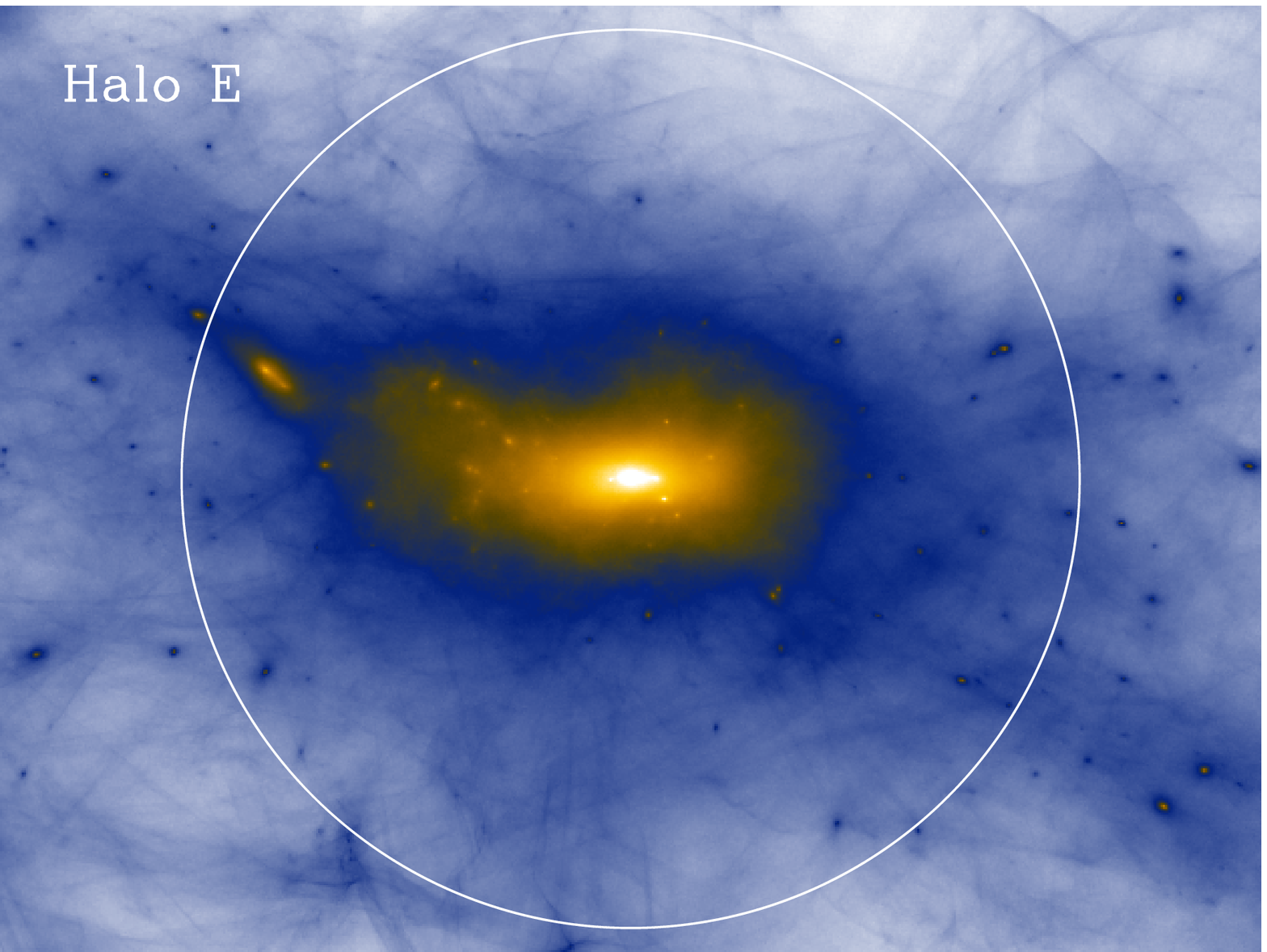} } 
\subfloat{\includegraphics[width=0.33\textwidth]{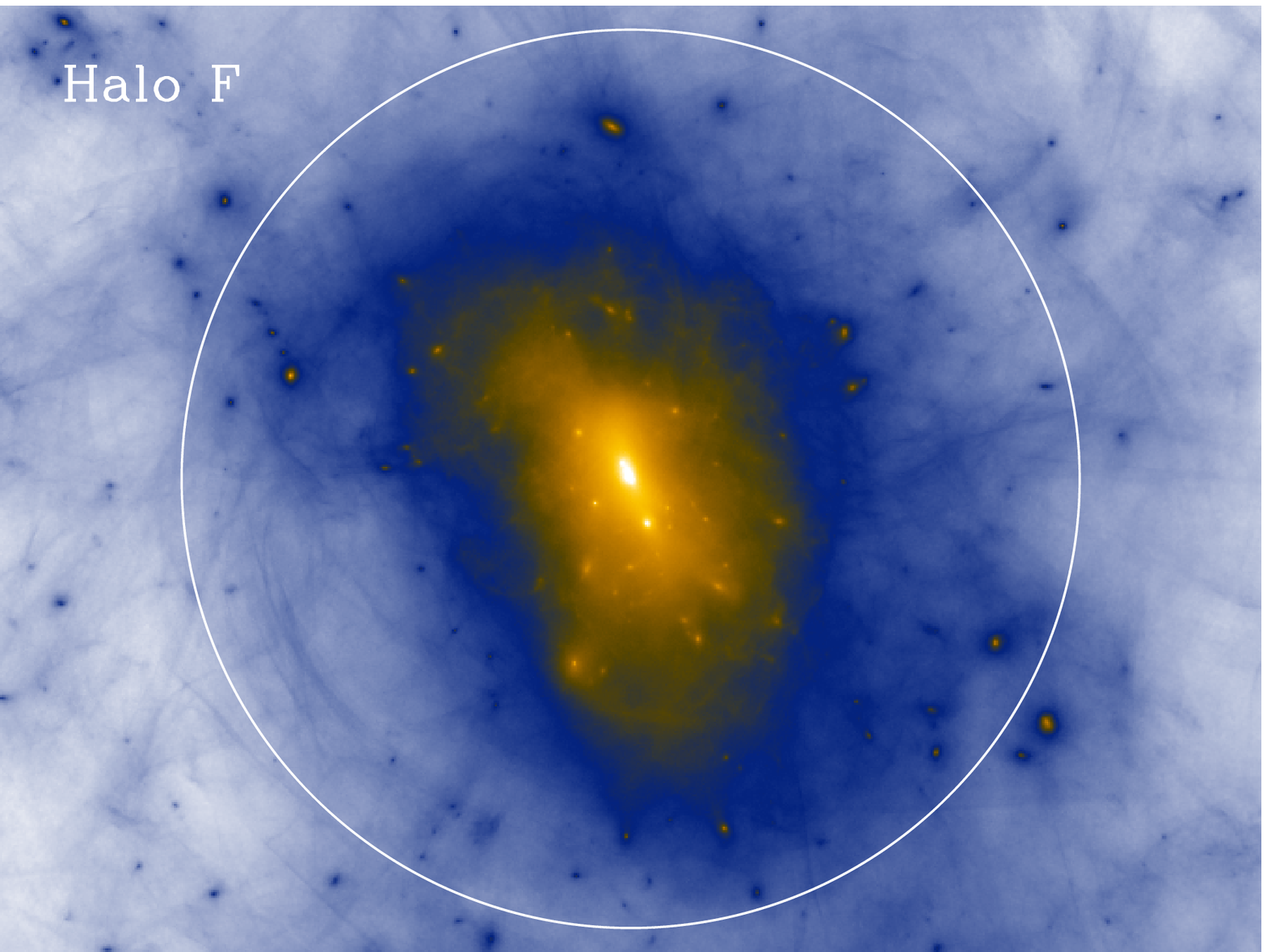} } 
\caption{Six of our resimulated dark matter haloes 
($\sim 10^8$ particles inside the virial radius), from the most massive
(top left) to the least massive (bottom right). The
image shows, using a logarithmic scale, the projected density of dark matter times
the square of the local dark matter velocity dispersion. The white circles
indicate the virial radius of each object. These images were constructed using
the method of \citet{Angulo2014a} based on a tessellation of the initial dark
matter sheet \citep{Abel2012,Shandarin2012,Hahn2016}, and are constructed
using an identical colour table in each panel \label{fig:zoom}}
\end{figure*}

\begin{table}
\begin{center}
\begin{tabular}{lcccc}
\hline 
Name $\qquad$& $M_{\rm vir}$         & $r_{\rm vir }$             & $N_{\rm vir}$ & $N_{\rm hr}$\\
     & $[h^{-1}\,\Msun]$ & $[h^{-1}\,{\rm pc}]$ &               & \\
\hline
\hline
Halo A & 0.0271 & 9.62 & $1.25\times10^8$ & $2.02\times10^8$ \\
Halo B & 0.0274 & 9.66 & $1.26\times10^8$ & $2.4\times10^8$ \\
Halo C & 0.0159 & 8.06 & $7.36\times10^7$ & $1.16\times10^7$ \\
Halo D & 0.0116 & 7.25 & $5.36\times10^7$ &  $8.8\times10^7$ \\
Halo E & 0.0127 & 7.48 & $5.88\times10^7$ &  $8.6\times10^7$ \\
Halo F & 0.0160 & 8.07 & $7.39\times10^7$ & $1.44\times10^8$ \\
\hline
\end{tabular}
\caption{Basic parameters of our high-resolution halo simulations at the final
output time, $z=30$. The free-streaming mass of the neutralino-DM model assumed
is $\sim10^{-6}\,{\rm M}_\odot$ and is thus resolved with $\sim10^4$ particles
of mass $2.163\times10^{-10}\,\Msun$.\label{tab:sims}}
\end{center}
\end{table}

At the $z=30$ output time, we identified haloes in the high-resolution parent subbox
using a FoF algorithm \citep{Davis1985} with a linking length $l=0.2$, and selected
for re-simulation the six 
most massive
objects, regardless of their properties. These haloes have virial masses of the order
of $10^{-2}\,\Msun$, and are resolved with between $10^4$ and $2\times10^4$
particles. Next, we selected a spherical region of radius $2\times{\Rvir}$
\footnote{Defined as the radius of a sphere with mean density equals to
$\Delta_{\rm vir} \, \rho_{crit}$, where $\rho_{crit}$ is the critical density of
the universe at a given redshift, and $\Delta_{\rm vir}$ is the overdensity of
virialized objects expected in the spherical collapse model \citep{Bryan1997}.
For the high redshifts we consider, $\Delta_{\rm vir}=\pi^2\simeq178$ is equal to
the virial overdensity for Einstein-de~Sitter cosmologies.}
surrounding each halo and tracked the positions of all enclosed particles back
to the unperturbed lattice. We then used {\sc Music} to identify the convex hull
containing these particles and to resample this region at higher
resolution.  Our re-simulations adopt a high-resolution particle mass of $\mps =
2.163\times10^{-10}\,\Msun$, which is equivalent to having sampled the full
simulation volume with $131\,072^3$ particles. Such high resolution allows us to
resolve the first non-linear objects (with masses $3\times10^{-6}\,\Msun$) with
roughly $10^4$ particles, and the final haloes with $\sim10^8$ particles. 

As for our parent run, the gravitational evolution was simulated using the {\tt
P-Gadget3} code, with forces computed using a Tree+PM algorithm. We employ two
particle meshes: one periodic mesh covering the whole simulation volume, and a
non-periodic one covering only the location of high-resolution region. Our runs
adopt a Plummer-equivalent softening length equal to $\epsilon=0.002$ comoving pc
(approximately $4000$ times smaller than the virial radius of our least massive
halo). Table~\ref{tab:sims} summarizes the main properties of
our re-simulated haloes. 

\begin{figure*} 
\centering
\subfloat[Halo A]{\includegraphics[width=0.16\textwidth]{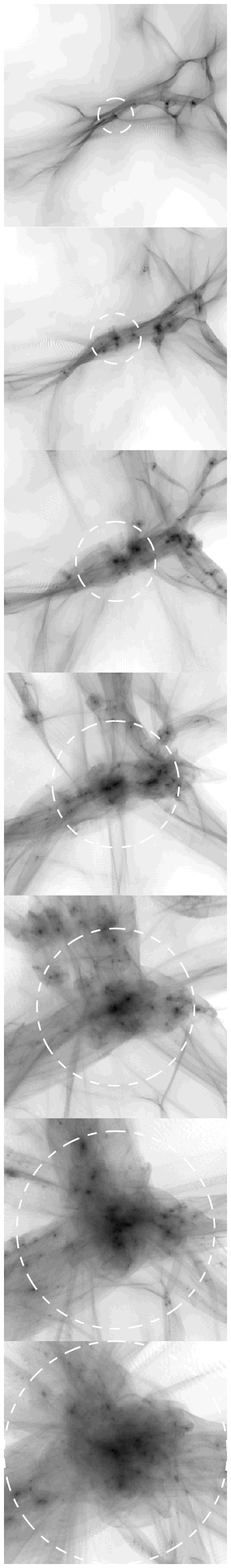} }
\subfloat[Halo B]{\includegraphics[width=0.16\textwidth]{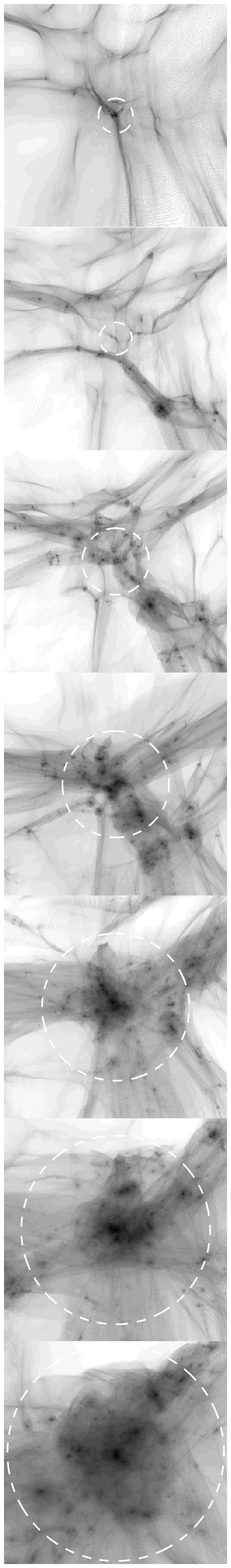} }
\subfloat[Halo C]{\includegraphics[width=0.16\textwidth]{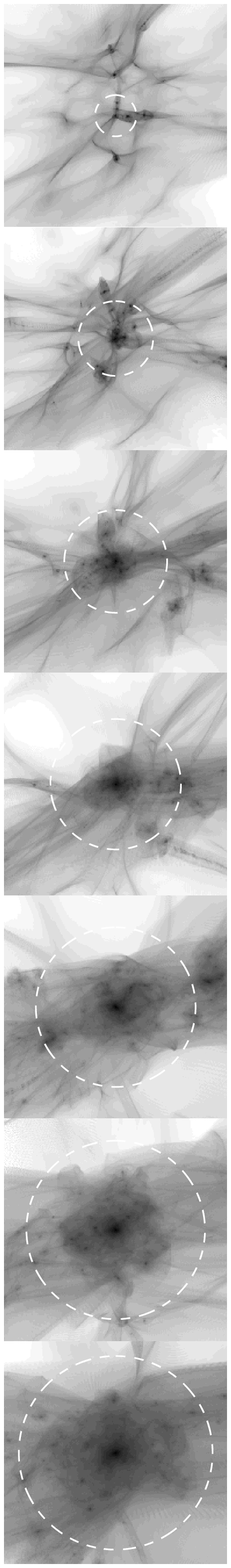} }
\subfloat[Halo D]{\includegraphics[width=0.16\textwidth]{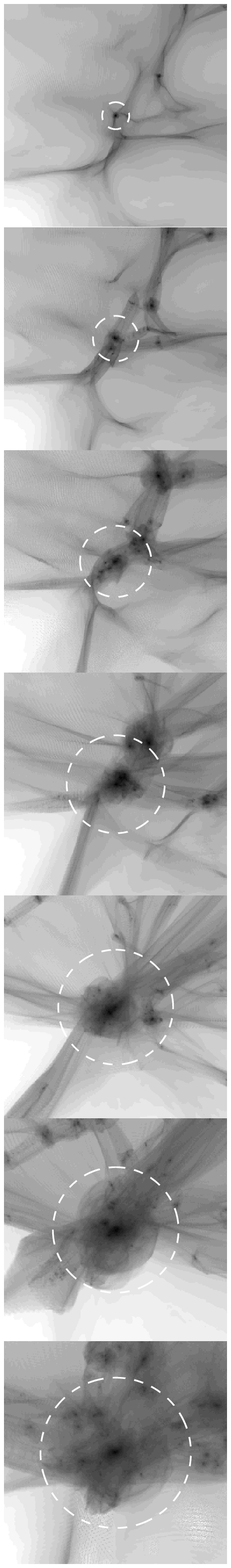} } 
\subfloat[Halo E]{\includegraphics[width=0.16\textwidth]{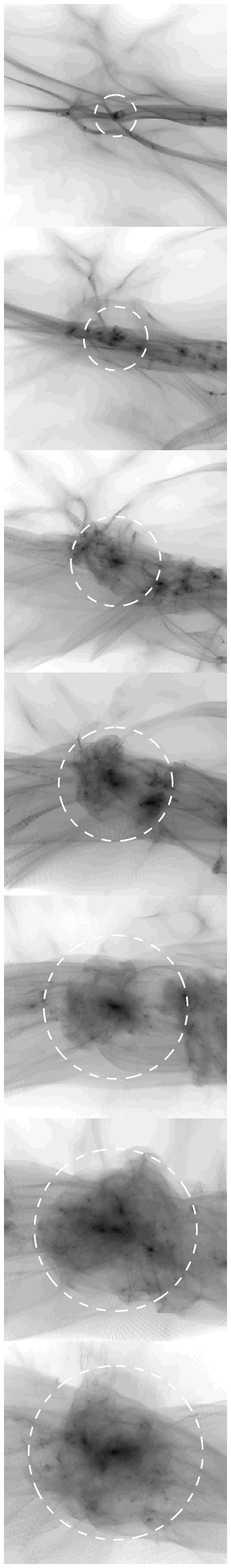} } 
\subfloat[Halo F]{\includegraphics[width=0.16\textwidth]{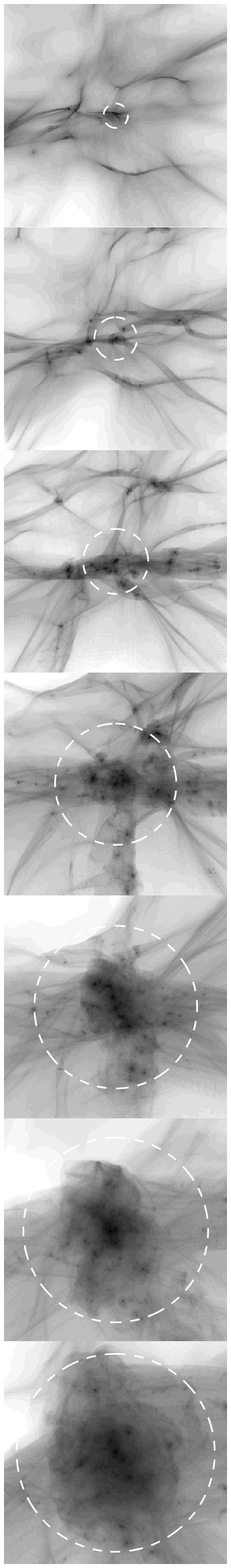} }
\caption{Snapshots of the time evolution of our six resimulated haloes (in
columns). From top to bottom we show the mass distribution inside a region of
$(12.5\,h^{-1}{\rm pc})^2$ with a projection depth of $10\,h^{-1}{\rm pc}$ at
redshifts $z=$ 57.8, 51.6, 46.6, 42.4, 39.0, 36.0, and 33.4 (top to bottom)
centred on the main progenitor. The circles indicate the virial radius of the
main progenitor at each output. The panels clearly show how the earliest
progenitors form from the collapse of smooth filaments, then proceed through a
phase of multiple rapid major mergers before they end up as relatively
isotropic haloes of $\sim10^{-2}\Msun$ at the latest times we consider.
\label{fig:evol}}
\end{figure*}

\begin{figure}
\includegraphics[width=8.5cm]{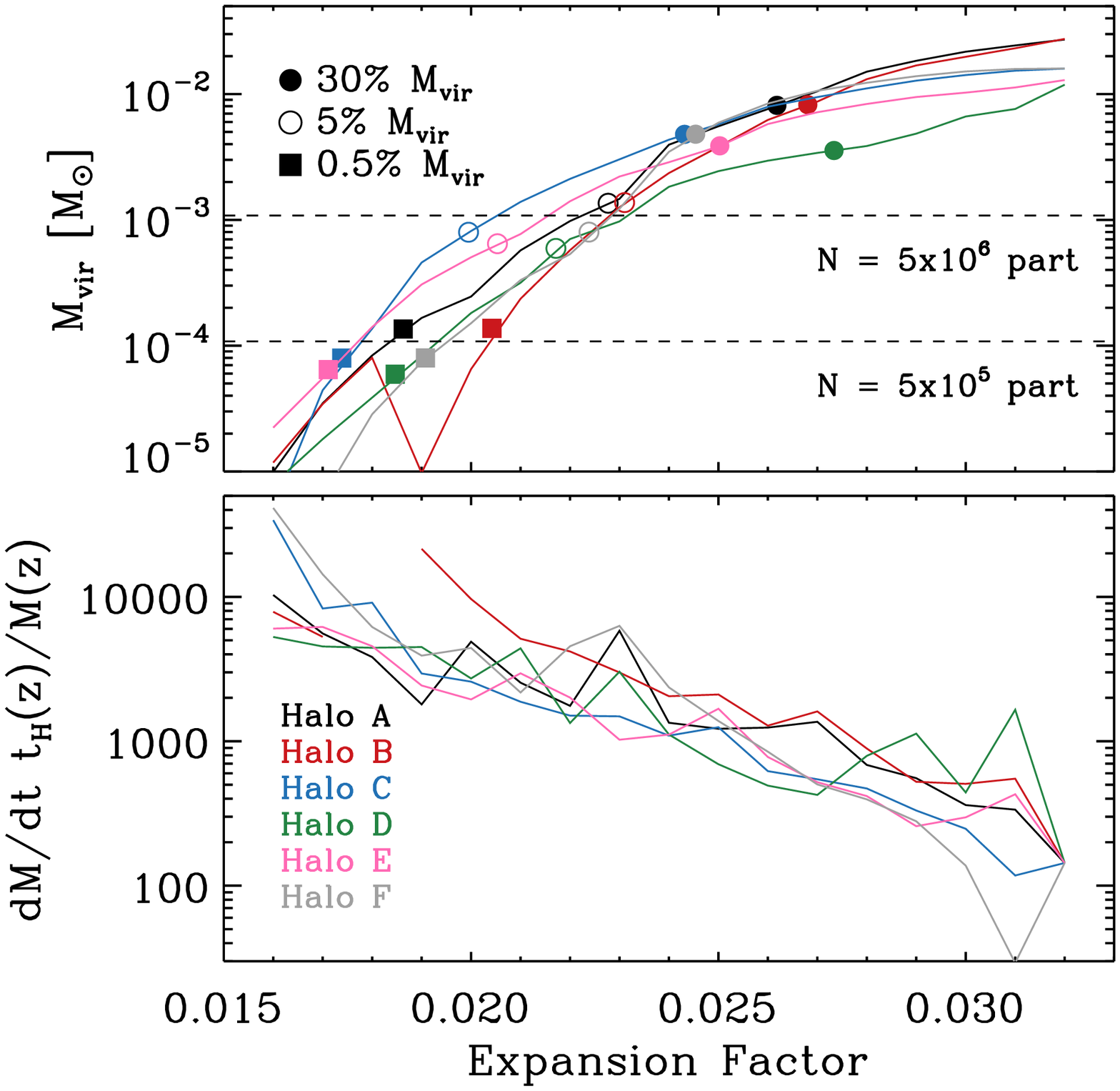}
\caption{ {\it Top}: The mass accretion histories of our 6 resimulated haloes. 
The dashed horizontal lines indicate the mass scale corresponding to $10^{6}$ 
and $10^{7}$ particles. Symbols denote the point along each MAH 
at which the main progenitor first reached $30\%$, $5\%$ and $0.5\%$ of the final 
halo mass. {\it Bottom}: Mass accretion rate, $dM/dt$, times the Hubble 
time, $t_H(z)$, divided by the current halo mass, $M(z)$. This quantity corresponds to 
the expected fractional increase in mass over a Hubble time, assuming a constant 
(instantaneous) accretion rate. \label{fig:growth}}
\end{figure}

\subsection{Visual Inspection}

Fig.~\ref{fig:zoom} shows the projected density distribution of dark matter at
$z=30$ for each of our $6$ resimulated haloes. Their internal structure is
well-resolved and shows an abundant population of substructure, signatures of
infalling filaments and of tidal disruption of satellite haloes. All of our
halos have well-defined centres and roughly spherical mass distributions. In
contrast, their high-redshift progenitors are much more perturbed, displaying
signatures of ongoing major mergers with several distinct structures of similar
mass.

The disturbed state of these haloes is not unexpected given their complex
merger histories. To illustrate this, in Fig.~\ref{fig:evol} we show images of
their main progenitors at seven characteristic times during their evolution,
(details of our centring and tracking algorithms are provided in Appendix A).
The columns in this figure correspond to the different haloes, and the rows to
the outputs at redshifts 57.8, 51.6, 46.6, 42.4, 39.0, 36.0, 33.4, from top to
bottom. In each panel, the virial radii of the main progenitors at this time
are marked with a white circle. Interestingly, at low masses, the virial radius
greatly overestimates the size (and shape) of the central collapsed region.
This is due to the presence of very dense filaments, which boost the
spherically averaged density in the regions surrounding the main progenitor,
and is not due to the presence of virialised material. Note that this also
implies that these objects are in fact much closer to the free-streaming mass
than their virial mass would suggest. 

The top rows, which corresponds to roughly when ${\rm M_{vir}}\approx
10^{-5}\,\Msun$, show that these haloes initially form at the centre or
intersection of a network of filaments.  This agrees with the classical picture
of halo formation in which a density fluctuation first collapses along one
axis, forming a pancake-like structure, then along a second axis forming a
filament which finally collapses onto itself leading to the formation and
virialisation of a halo \citep[see also][for a similar picture in
WDM]{Angulo2013}. Also evident in each of the top panels are several additional
haloes of similar size to the main progenitor, but a dearth of lower-mass
systems. 

By the time the main progenitors have reached 10\% of their final mass, they
have undergone a series of major mergers with nearby structures. Notable
examples are Halo B, which has of order 20 density peaks of comparable mass
within its virial radius, and Halo E, which has a late major merger. As their
mass continues to grow, our halos start to resemble typical CDM haloes; all of
them have well-defined centres, approximately spherical mass distributions, and
an abundant population low-mass substructure. By the final time, they appear to
be in a relatively relaxed state.  

The very rapid mass growth of our haloes is highlighted also in
Fig.~\ref{fig:growth}, where we show the mass accretion histories (top panel)
and accretion rates (bottom panel) for all 6 haloes. These haloes increase
their mass by roughly three orders of magnitude in less than $60$ Myrs, a
fractional accretion rate that is millions of times higher than that of cluster
sized haloes at $z=0$ \citep{Fakhouri2010}. 

\subsection{Density Profiles}

\begin{figure}
\includegraphics[width=0.5\textwidth]{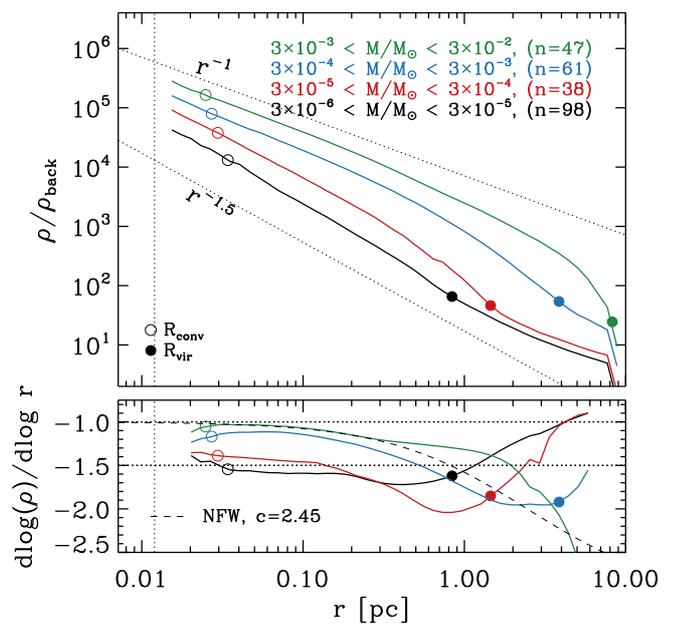}
\caption{ The average density profiles (top panel) and their logarithmic slopes
(bottom panel) in bins of mass for all haloes identified (at all times) in the
high-resolution region of our simulations. The range of halo masses included in
each bin, as well as the number of objects in each, is indicated in the legend.
The open and filled circles along each curve indicate the average convergence
and virial radii, respectively. The vertical dotted line marks the formal
resolution limit of our simulations, i.e. $2.8\times \epsilon$.  For
comparison, the dashed line in the lower panel shows the logarithmic slope of
an NFW profile with concentration $c=2.5$. \label{fig:bino_stacked} Open
circles indicate the minimum scale for which we expect converged results,
as estimated by the algorithm described in Power et al. (2003).
}
\end{figure}

\begin{figure*}
\includegraphics[width=0.95\textwidth]{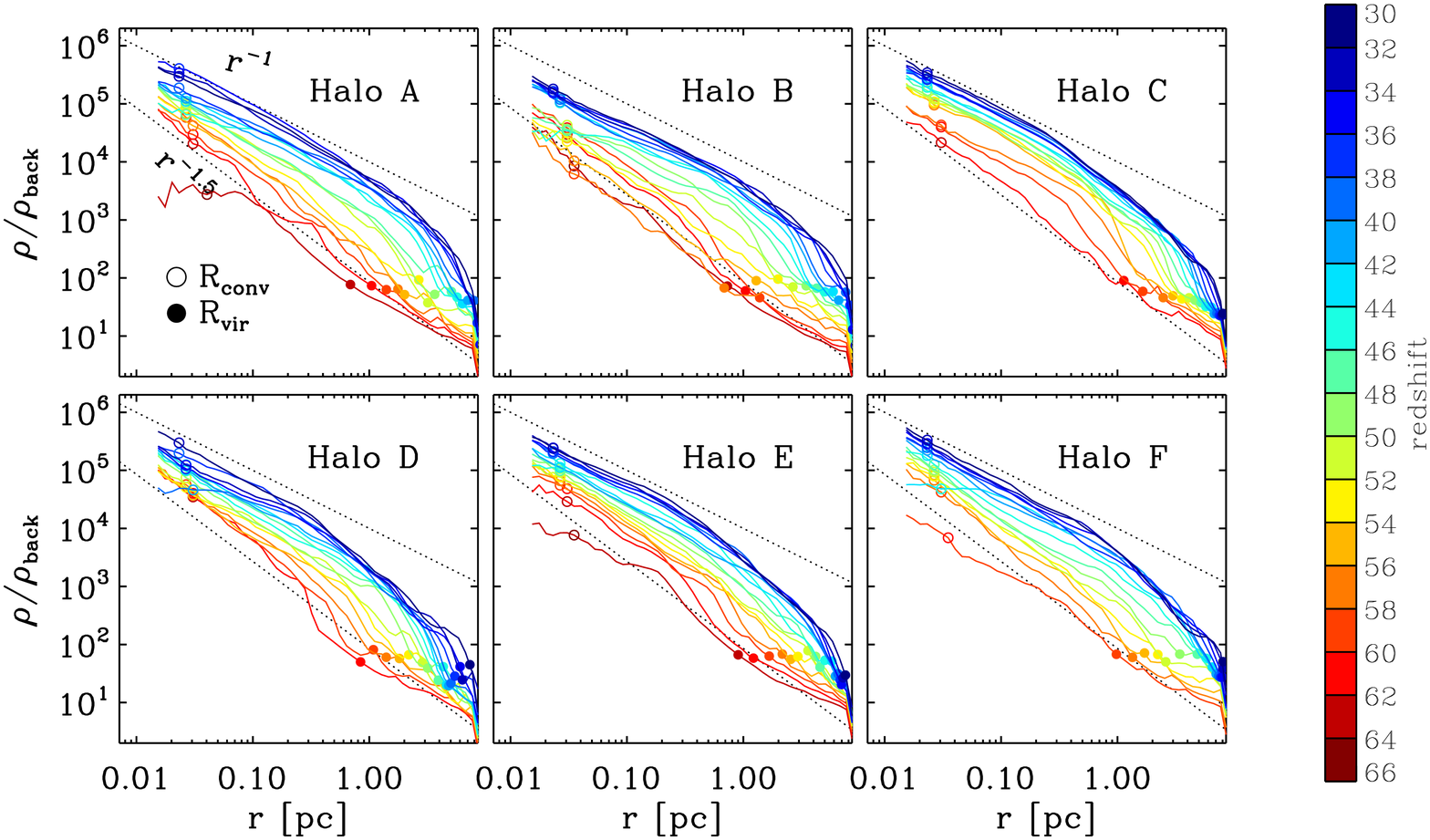}
\caption{The evolution of the spherically averaged density profiles for the
main progenitor of each of our 6 haloes. Different colours show results for
different redshifts, as indicated by the colour bar on the right-hand side.
Densities are normalized by the background density at each redshift, and radii
are expressed as the comoving distance to the halo centre. The dotted lines
highlight two power-law profiles, $\rho\propto r^{-1.5}$ and $r^{-1}$. Filled
circles along each curve indicate the virial radius of the main progenitor and
open circles the convergence radius \citep[see][]{Power2003}.
\label{fig:bino_profiles}}
\end{figure*}

\begin{figure*}
\includegraphics[width=0.95\textwidth]{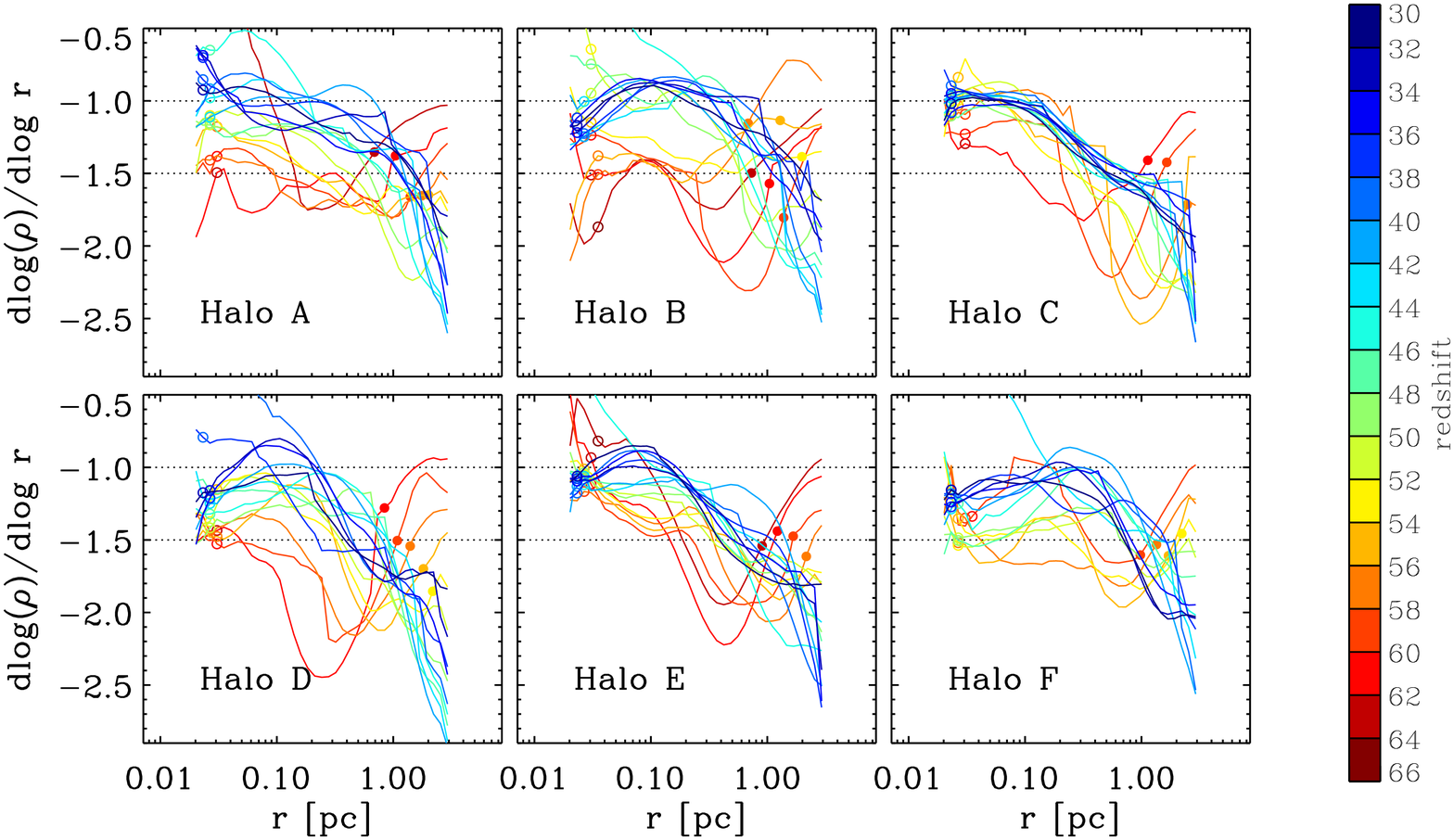}
\caption{Same as Fig.~ \ref{fig:bino_profiles} but showing the logarithmic
slope of the density profiles. 
\label{fig:bino_slopes}}
\end{figure*}

Before exploring the density profiles of our individual haloes, in
Fig.~\ref{fig:bino_stacked}, we show the {\em average} profiles in bins of halo
mass across all output times. To construct these profiles, we first identified,
in each simulation output, all haloes with $\Mvir > 3\times10^{-6}\,\Msun$
(corresponding to $\sim10^{4}$ particles) existing in the high resolution
region of our simulations. We then compute the spherically averaged density
profile for each of these haloes by summing simulation particles surrounding
the halo centre in equally spaced logarithmic bins of width $\Delta \log r =
0.2$ from $0.001\,$pc to $10\,$pc. Finally, we add the spherically averaged
density profiles after separating them into four disjoint mass bins. This
substantially reduces the noise in our density estimates (each mass bin
contains at least 38 haloes).

At the smallest masses, just above the free-streaming scale, the inner density
profile follows closely a power-law with a slope close to $-1.5$ (the lower
diagonal dotted line in the top panel shows a pure $\rho\propto r^{-1.5}$
power-law). On intermediate scales the profile exhibits a sharp decline,
marking the position of the outermost caustic and highlighting the size of the
collapsed region \citep[cf. also][]{Diemer2014}. Note that this scale, in
agreement with our expectations from Fig.~\ref{fig:evol}, is $3$-$4$ times
smaller than its virial radius (denoted by filled circles). As discussed above,
the classical ``virial'' radius overestimates the true isotropised region of
the halo at the earliest times after collapse. On larger scales, extending
beyond the virial radii, the density profiles become shallower, possibly
reflecting the density profile of the Lagrangian ``proto-halo''. 

As halo mass increases, there is a systematic change in the shape of the
density profile. On small scales, the inner slope becomes shallower,
approaching $-1$ (the NFW value) for the largest mass bin considered.
Quantitatively, the measured halo mass dependence of the inner slope is in good
agreement with the results of \cite{Ishiyama2014}. The values measured from the
average density profiles in our simulations are [-1.0, -1.1, -1.4, -1.6]
respectively for the higher to the lower mass bins considered, whereas the
best-fit relation reported by Ishiyama et al. yields [-0.97, -1.1, -1.22, -1.34].
These results might either suggest that the NFW profile is a dynamical
attractor, or simply that haloes of different mass form in fundamentally
differently ways (e.g. a from protohaloes with different shapes and/or density
profiles). We will investigate this next.

To explore these issues in more detail, we show the density profiles of our six
re-simulated haloes in Fig.~\ref{fig:bino_profiles}. In each panel, the
sequence of coloured curves shows the evolution of the spherically-averaged
density along the main progenitor branch; red curves correspond to the highest
redshifts, $z\approx 70$, and blue to redshifts closer to $z\approx 30$, the
final output.  The corresponding logarithmic slopes are shown in
Fig.~\ref{fig:bino_slopes} for the same output times (note that the slope at
radius $r_i$ is estimated from the best power-law fit to the density measured
in six adjacent radial bins). 

Consistent with the picture discussed above, as our re-simulated haloes grow in
mass, both the {\em normalisation and the shape} of their spherically-averaged
density profiles show clear evolution with time. This implies that the profile
transformation occurs on an individual halo basis rather than being a
consequence of haloes of different masses being governed by different collapse
and/or virialisation mechanisms. Based on this, and on the images shown in
Fig.~\ref{fig:evol}, we may summarise a few points of interest: 

1) At the earliest times, just after the haloes emerge from the monolithic
collapse of a filament, the halo's inner density profile is well-described by a
power-law with a slope close to $-1.5$ (note that the exact asymptotic slope
cannot be robustly determined in every case due to the intrinsic scatter in the
measured profiles).

2) All haloes evolve through a period of primarily smooth mass accretion: the
outermost caustic moves closer to the virial radius, and the central density
profile changes very little in either amplitude or slope. For a short time, all
profiles remain close to a steep single power-law.

3) For all haloes we observe a change in the inner density slopes, which
typically changes from $\sim-1.5$ to $\sim -1.0$ over the redshift range
$z\sim60-40$. This phase appears to coincide with the collapse and accretion of
several other haloes of similar mass, resulting in a phase of approximately
simultaneous major mergers. After this period of rapid change, the functional
form of the profiles appears more stable again.  It is important to note that
the final density profiles can be {\em shallower} than $-1$ in some cases.

In summary, our results suggest that the density profiles of the descendants of
the first neutralino haloes are a consequence of the particular details of
their merger history, rather than only being a manifestation of their
Lagrangian peak shapes. Additionally, these results also suggest that properly
resolving the first haloes and their initial merger history is necessary in
order to predict the properties of more massive systems forming at later times.
To elucidate the transformative effects of major mergers in shaping the density
profiles of DM haloes, in the next section we explore the impact of a sequence
of approximately equal mass, rapid mergers on an initially steep power-law
profile. 

\section{Controlled Simulations}

As we have demonstrated in the previous section, $\chi$DM microhaloes are born
with single power-law profiles and slowly transition to profiles resembling the
well-known NFW form during phases of rapid mergers. We next investigate this
transformation process in non-cosmological simulations in detail. Specifically,
we now consider the impact of isolated and repeated mergers on an initially
cuspy, spherically-symmetric dark matter halo with an isotropic velocity
distribution.

\subsection{Numerical Methods}
We first summarise how we set up our idealised simulations of mergers of
non-cosmological haloes. Specifically, the density profiles we assume, how we
sample them with N-body particles, and how we set up the mergers of several of
such haloes.

\subsubsection{Initial conditions}
\label{sec:dmhalomodels}

For our simplified halo models, we assume an initial power-law density profile, 
truncated at large radii, of the form

\begin{equation}
\label{eq:rho_isolated}
\rho(r) = \rho_s \left(\frac{r}{r_s}\right)^{-\gamma} \erfc[\kappa (r/r_s - 1)].
\end{equation}
Here $\rho_s$ and $r_s$ are the arbitrary scaling variables of density and radius, 
respectively, and $\kappa$ defines the cut-off scale.
We consider three power-law indices, $\gamma = 1.0$, $1.5$, and $1.9$, and in
all cases adopt $\kappa = 2$ and $r_s = R_{vir}$; $\rho_s$ is chosen to 
fix the halo mass to the desired value. In addition to
these scale-free models, we also consider haloes with a Hernquist density
profile \citep{Hernquist1990} with a concentration parameter of $R_{vir}/r_s =2.5$. 
We show the distribution functions associated with each of these four
cases in Figure~\ref{fig:df} of Appendix~\ref{sec:setting_up_profiles}.

Each of our models samples the density distribution using $10^6$
particles\footnote{Note that this refers to the total particle number within
the virial radius. Because the mass model extends to slightly larger radii, the
total number of particles used for each halo is slightly larger.}. Velocity
magnitudes, $v$, are drawn directly from an isotropic distribution function,
which we calculate using Eddington's formula \citep[e.g.][]{Binney2008}: 
\begin{equation}                                                                                              
\label{eq:df}                                                                                                 
f(E) = \frac{1}{\sqrt{8} \pi^2} \left[ \int_0^E \frac{{\dd}^2\rho}{{\dd}\Psi^2}                               
       \frac{{\dd}\Psi}{\sqrt{E-\Psi}} + \frac{1}{\sqrt{E}} \left(\frac{{\dd}\rho}{{\dd}\Psi}\right)_{\Psi=0}\right].
\end{equation}
Here $E \equiv -\Phi(r)-\frac{1}{2}v^2$ is the relative energy and $\Psi \equiv
-\Phi$ the relative potential. Particle radii and
velocities are then assigned random orientations such that the resulting halo
is isotropic in both position and velocity space. We note that this procedure
yields more accurate and robust results than the widely-used local Maxwellian
approximation \citep[see also][]{Kazantzidis2004}. We describe our procedure 
for setting up these initial conditions in more detail in Appendix~\ref{sec:setting_up_profiles}, 
and test the stability of our model haloes in Appendix~\ref{sec:stability}. 

\begin{figure}
\includegraphics[width=0.45\textwidth]{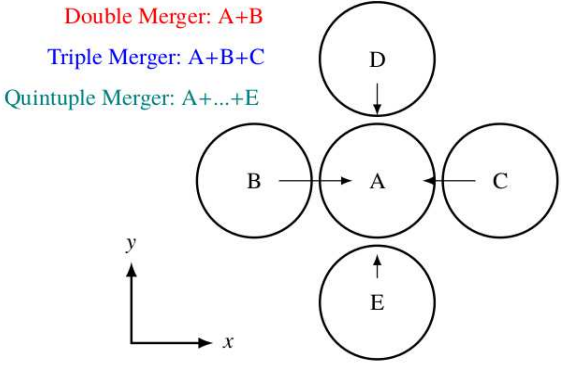}
\caption{Diagram showing the possible initial configurations of our merger simulations. All
haloes are initially placed on the x-y plane, and are displaced by two 
virial radii from the central halo A.  Arrows indicate the magnitude and
direction of the initial velocity vector of the haloes: all are set on 
radial orbits with velocities equal to 100\%, 75\%, 50\%, and 25\% of the
escape speed measured at the virial radius of halo A.}  
\label{fig:sketch}
\end{figure}

\subsubsection{Model parameters and initial setup of the mergers}

To explore the impact of multiple major mergers on the internal density profile
of a halo, we simulate a sequence of head-on mergers of identical, spherically
symmetric haloes generated using the procedure described above.
Fig.~\ref{fig:sketch} shows the merger configurations, of which we will
consider 3 cases: 1) a binary merger between haloes A and B, 2) a triple merger
between haloes A, B and C, and 3) a quintuple merger between all five haloes.
Initially, the centres of all haloes lie on the same plane and are separated
from halo A by $2 \times R_{vir}$.

We choose a coordinate system so that halo A is initially at rest, and set the
remaining haloes on radial orbits with relative velocities of $v_{\rm
rel}=-v_{\rm esc}$, $-3/4\,v_{\rm rel}$, $-1/2\,v_{\rm rel}$ and $-1/4\,v_{\rm
rel}$ for haloes B, C, D and E, respectively. Here $v_{\rm esc}$ is the nominal
escape speed measured at the virial radius of Halo A. The differences in the
relative velocities of the haloes ensure that the mergers are not
synchronized, but also that each occurs before the system has had time to
dynamically recover from prior mergers. Although other configurations are
clearly possible, this choice is motivated by the very violent merger histories
of our neutralino haloes with near-radial orbits.

All merger simulations were carried out using {\tt P-Gadget3}, the same code
that was used for our cosmological runs. Forces were computed using only the Tree
algorithm, with open boundary conditions within a static (non-expanding)
background. The softening length was set to $\epsilon = 0.3$ kpc,
approximately $0.2\%$ of the initial virial radius of each halo. Forces are
therefore exactly Newtonian for separations larger than $2.8\times\epsilon =
0.84$ kpc. All simulations were evolved for $15$ Gyrs ($\sim8$ dynamical
times), and employed approximately $9000$ adaptive time steps. We output the
particle data at 16 different times, with the last five separated by $0.01$
Gyrs. To reduce noise, the final results of these simulation will be presented
as an average over those five final output times. All models are initialised to 
have a virial mass of $10^{12}\,{\rm M}_{\odot}$ and a virial radius of $162.2$ kpc. 
Although this is a very different mass scale than that of our neutralino haloes,
the scale-free nature of gravity guarantees that this choice is irrelevant.

\subsection{Merger Remnants for $\gamma=1.5$ profiles}

\begin{figure}
\includegraphics[width=0.5\textwidth]{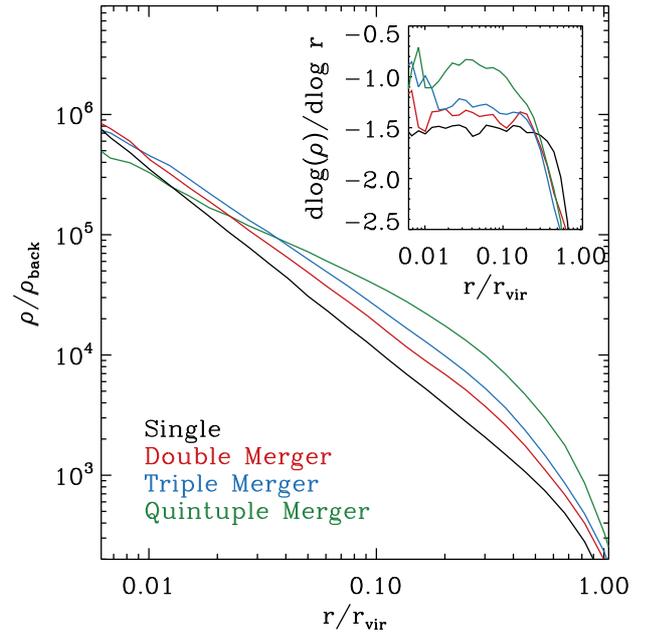}
\caption{The density profiles, in units of the background density, of the final
remnants of equal-mass mergers between haloes with initial power-law density
profiles with $\gamma=1.5$, c.f. eq.~(\ref{eq:rho_isolated}). Different colours show the
results for an isolated halo (black), and for two, three and five mergers (red,
blue, and green, respectively). The inset shows the logarithmic density slopes.
In both plots, the x-axis has been scaled by the initial virial radius of 
halo A. \label{fig:gamma}}
\end{figure}

\begin{figure}
\includegraphics[width=0.5\textwidth]{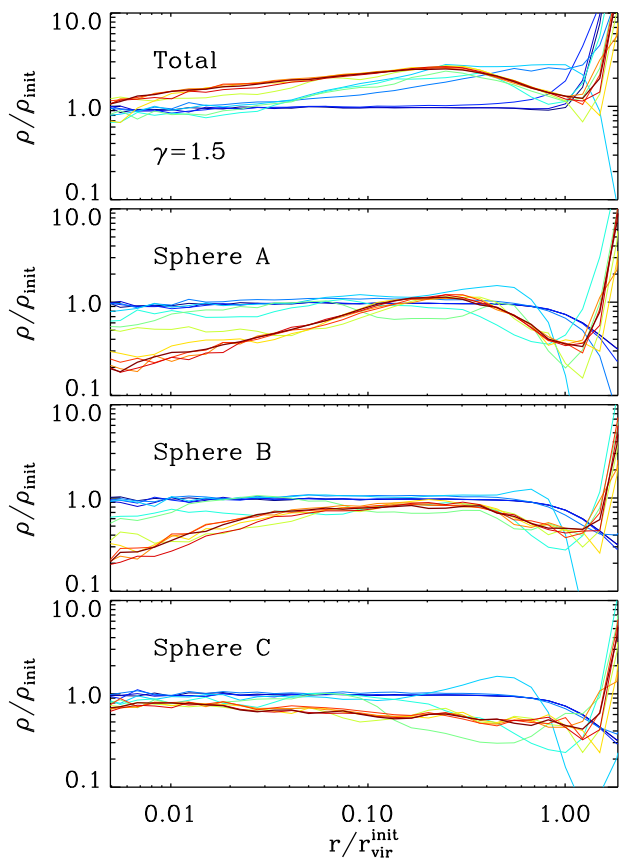}
\caption{The final density profile, in units of the initial value, of the 
triple $\gamma=1.5$ merger simulation. The top plot shows the results for all particles,
whereas each of the lower panels focuses on the mass contributed by each of
the three progenitors, as indicated by the legend. In 
each panel, coloured lines indicate the results at different output times;
dark blue corresponds to the initial times, whereas dark red corresponds
to the final times. \label{fig:sep_profile}}
\end{figure}

We first study the outcome of the merger of up to five such idealised haloes
that are initially given by the truncated single power-law profiles given in
eq.~(\ref{eq:rho_isolated}) with a slope of $\gamma=1.5$. Specifically, we investigate 
how their density profiles are changed as a result of the mergers and focus on
how the inner asymptotic slope evolves away from the initial value of $\gamma$.

In Fig.~\ref{fig:gamma} we show the remnant density profiles for mergers 
between initially isotropic haloes with $\gamma=1.5$ power-law density cusps
(the initial density profiles in this case are a good match to those of the first
generation of neutralino haloes). Lines of different colour correspond to the 
different merger configurations discussed above.

The most striking result seen here is that the final profile departs
significantly from its original power-law form, and exhibits an inner slope
that becomes systematically shallower as the number of mergers increases. This
is best appreciated in the inset plot, which displays the logarithmic slopes of
the profiles. For the binary merger, the initial power-law index of $-1.5$ has
decreased slightly, to $\sim -1.4$, whereas for the quintuple merger it has
decreased to $\sim -0.8$, a value much closer to (and even lower than) the
asymptotic slope of the NFW profile. At the same time, the outer slope becomes
steeper. Overall, the homology of the original density profile is broken and
its final shape depends sensitively on the particular details of its merger
history. This is qualitatively very different than the outcome of mergers
between NFW haloes, which primarily affect the profile's concentration, but not
its shape \citep[e.g.][]{Kazantzidis2006}. In addition, there is no single
asymptotic value for the inner slope, which apparently disfavours the
interpretation that the NFW profile is a natural outcome of major mergers.

We now take a closer look at the triple merger case by following the time evolution
of the density profile, separately considering the particles initially associated with
each progenitor. Fig.\ref{fig:sep_profile} shows the spherically averaged density
profiles after normalizing by the initial analytic profile. Different panels,
from top to bottom, show results for all particles, as well as for particles belonging 
to Haloes A, B, and C, respectively. For each particle subset, and at every output time, 
we have recomputed the centre-of-mass before calculating $\rho(r)$.

We can see clearly that the final density profile is not simply a rescaled
version of the initial one, reinforcing the results discussed above.
Furthermore, the distribution of particles that originated from each progenitor
also changes. For Haloes A and B, the first two to collide, the central
densities decrease significantly and rapidly, by about a factor of 5 in less than 
$\Delta t = 1\,{\rm Gyr}$. The outer parts also suffer a major decrease in density. 
Because the density does not increase above the
initial value at any radius, the removed particles have necessarily moved
beyond the initial halo boundaries (causing the quantity plotted to reach
values above unity for $r>r_{vir}$). On the other hand, particles from Halo C
appear to have suffered a qualitatively different transformation: densities
decrease on all scales, but only slightly so. In fact, particles from Halo C
dominate the central regions of the final halo. We note that this, however, is
not a universal behaviour. For other merger configurations, Haloes A or B end
up dominating the central regions, while the last sphere accreted is more
strongly disrupted. Thus, the transformation suffered
by a particular sphere seems to depend on the particular details of the 
merger. What seems ubiquitous however, is the emergence of an inflection point
in the total profile: a scale above and below which the distribution of
densities decreases. In the following section we will explore the origin of
these transformations in more detail.

\subsection{Energy distribution and Violent Relaxation}

During major mergers, the relative motion of large amounts of mass leads to
large temporal variations in the gravitational potential. In a process known as
violent relaxation \citep{Lynden-Bell1967}, these potential fluctuations change
the energies of individual particles, resulting in a final equilibrium state
whose mass distribution can be substantially different than the initial one.
We argue that it is indeed violent relaxation that drives the evolution of
the density profile of the merger remnant. In this section, we provide evidence 
for this scenario in terms of the evolution of the particle energy distribution
function as well as the temporal evolution of the central potential during
the merger.

\begin{figure}
\includegraphics[width=8.5cm]{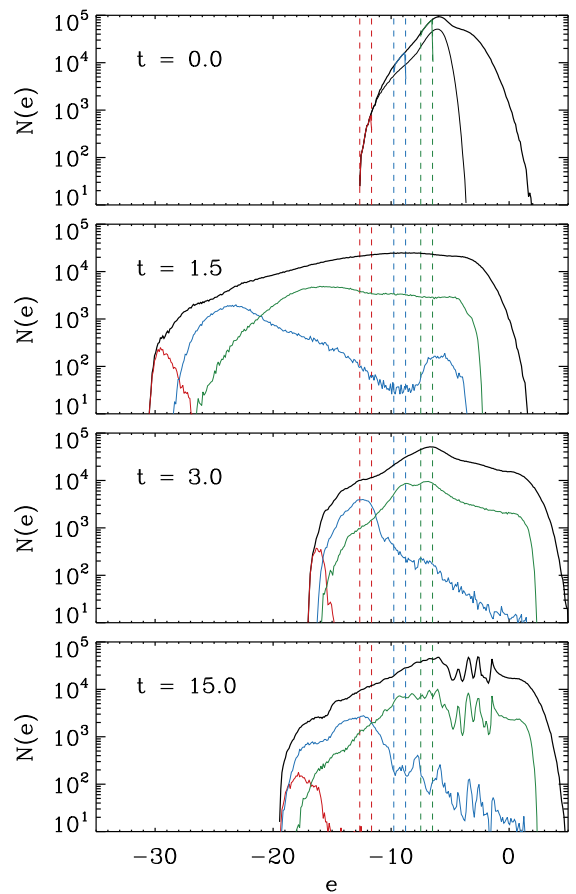}
\caption{Time evolution of the distribution of energies of particles in the
triple merger of $\gamma=1.5$ spheres. The top panel shows the initial
configuration. The second panel, $t=1.5$, corresponds to the time when the
potential reaches its minimum value. The next panel, $t=3.0$, corresponds to
when the potential reaches a local maximum, as the merger remnants recede. The
black line shows the distribution for all particles involved in the mergers,
whereas coloured lines follow the distribution of only those particles inside
three initially narrow energy bins, indicated by the vertical dashed lines of
the same colour.
\label{fig:time-energy}}
\end{figure}

\begin{figure}
\includegraphics[width=8.5cm]{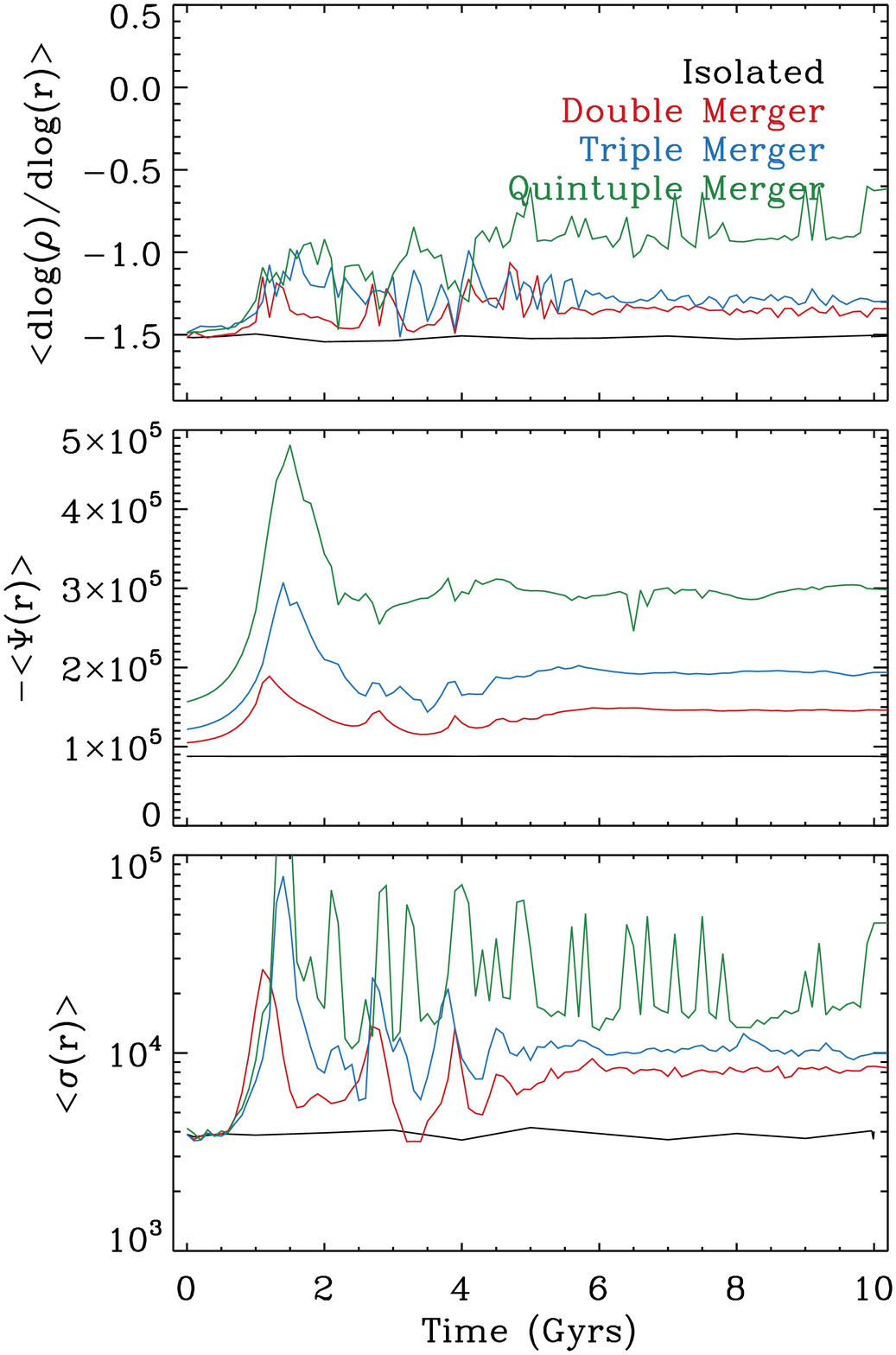}
\caption{The time evolution of the average density profile slope (top panel),
gravitational potential (middle panel), and velocity dispersion (bottom panel)
of merger remnants of halo models with $\gamma=1.5$. The average density
slope is measured by fitting a power-law over the range $0.01 < r/\Rvir < 0.1$
to each measure density profile, while the average potential and velocity
dispersion were computed over a smaller radial range, $0.01 < r/\Rvir < 0.05$.
\label{fig:pot} }
\end{figure}

\begin{figure*}
\includegraphics[width=18cm]{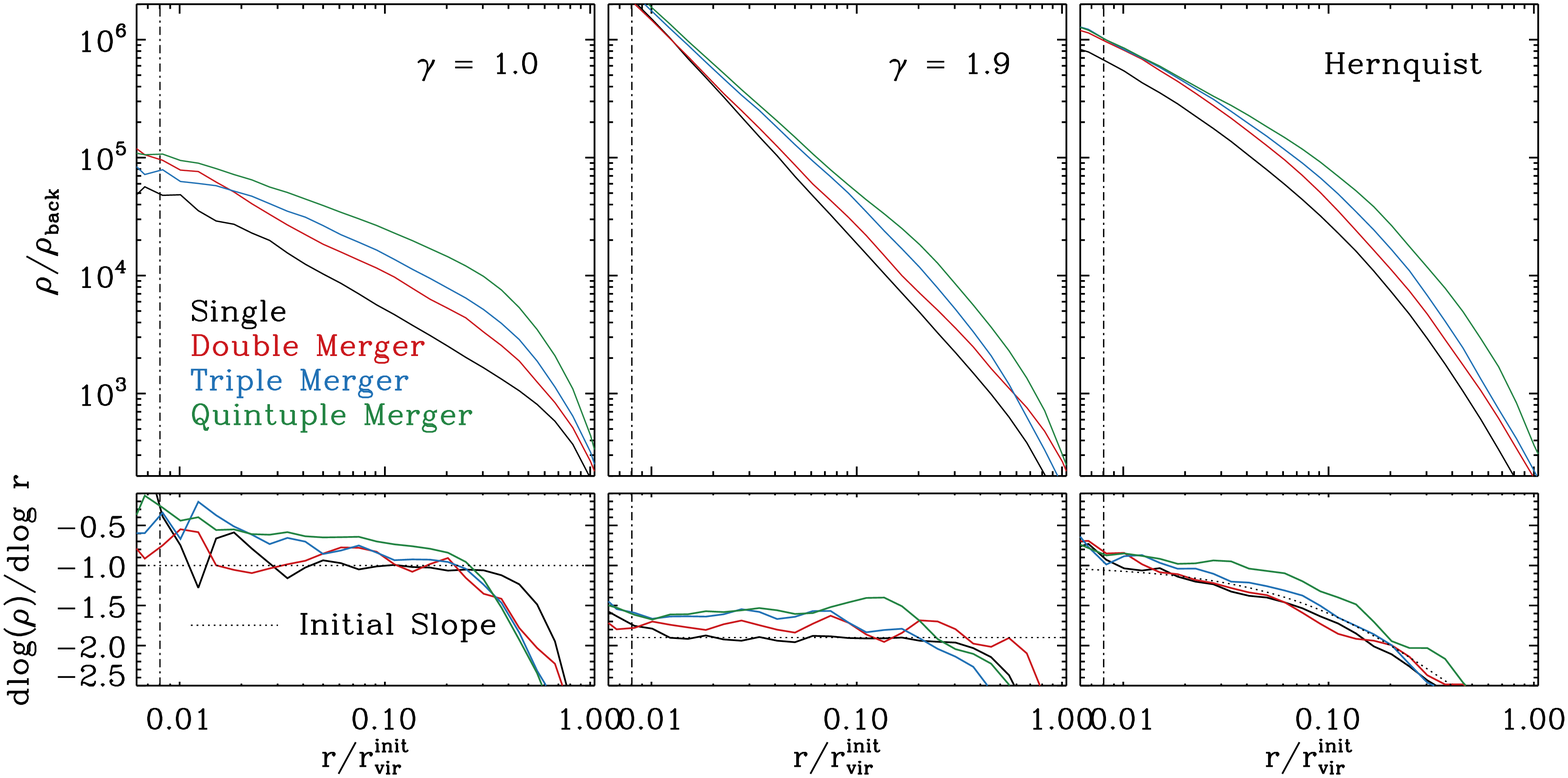}
\caption{The final density profiles for our (non-cosmological) simulations of
equal-mass mergers. The top panels show the spherically averaged density
in units of the background density; bottom panels show the corresponding
logarithmic slopes. Each column displays results for models with different
initial density profiles, with different colours corresponding to different 
merger scenarios, as indicated in the legend. The ordinate plots the radial variable
in units of the initial virial radius of each halo.
\label{fig:spheres}}
\end{figure*}

In Fig.~\ref{fig:time-energy} we plot the time evolution of distribution of energies for
particles involved in the triple merger. The initial distributions for the
whole system and for the central sphere are shown in the top panel using thick
and thin solid black lines, respectively.  Coloured lines follow the energy
distributions of particles inside three narrow bins of initial energy, marked
in each panel using vertical dashed lines of similar colour. Green lines follow
an initially weakly bound group of particles, blue lines correspond to
particles with average energy, and red lines to the most bound particles.

As expected for a system undergoing violent relaxation, the initially narrow
energy distributions broaden significantly as a result of the mergers. During
these events, the potential energy fluctuates (as we will explicitly show
below) but the total energy must remain constant; the system responds by
increasing or decreasing the kinetic energy of individual particles as it
settles towards a new equilibrium state. This oscillatory behaviour modifies the
energies of individual particles thereby broadening the overall  distributions.
The most bound particles become even more tightly bound, whereas many weakly
bound particles gain enough kinetic energy to become unbound. Although not
shown here, the broadening of the energy distributions is more substantial for
mergers involving a larger number of progenitors, which drive larger
fluctuations in the underlying gravitational potential of the system.

At $t=1.5$ (which coincides with the minimum value of the potential energy),
the energy distributions become very broad, and exhibit a much higher absolute
mean. The increased depth of the gravitational potential offers new energy
states, making it possible for particles to become more tightly bound. At
$t=3$, when the potential has decreased substantially, the average and the
dispersion in the energy distributions also decrease, as the merger remnants
are partially destroyed and no longer contribute to large potential
fluctuations. At this point, some particles reach positive energies and escape
the system entirely. By the final time, the velocity dispersion of particles
stabilizes but, due to the increased mass of the system as a whole, the binding
energy of all three groups has increased (i.e. become more negative).
Furthermore, the initially smooth distribution develops oscillatory features at
low energies, which are imprinted by the particular merger configuration. 

These results suggest that violent relaxation can drive large changes in mass
profiles of haloes, ultimately transforming their density profiles from one
form to another. To highlight this, in Fig.~\ref{fig:pot} we plot the time
evolution of the logarithmic density slope (top panel), the central
gravitational potential (middle panel), and of the central velocity dispersion
(bottom panel). The later two quantities were computed in the range  $0.01 <
r/\Rvir < 0.05$. Differently coloured lines show the results for the evolution
of an isolated halo (black), and for three merger configurations considered, as
indicated in the legend. 

We can see that while the isolated halo shows an invariant slope and potential
at all times, the various mergers display strong variations in all quantities.
For all configurations there are several rapid increases in $\Psi$ (by factors
of 3 to 8, depending on the number of systems involved in the merger) as the
merging systems approach one another. $\Psi$ reaches its maximum during
pericentric passages, and decreases as the merger remnants recede. The
amplitude of fluctuations in the potential is rapidly suppressed in subsequent
pericentric passages, owing to phase-mixing (violent relaxation is
self-limiting). The whole system then approaches a new equilibrium and the
gravitational potential stabilizes at a value in between the initial and
maximum values. We can see that variations in the potential coincide with
variations in the specific kinetic energy (or $\sigma$) as well as in the
density slope. This supports our previous conjecture that major mergers are
directly responsible for the transformation of the density profile observed in
our cosmological simulations.

\subsection{Merger remnants for other density slopes}

We saw above that violent relaxation can transform initially steep density
cusps into substantially shallower ones which provides a natural explanation
for why the initial $r^{-1.5}$ profiles should not survive through phases of
multiple major mergers. In the cosmological simulations of the first part of
the paper, the profiles approached a significantly shallower inner slope and in
particular the broken power-law profiles of the NFW type during later stages of
evolution. A remaining open question is thus whether there is a natural
end-point to the evolution of the density profiles or whether slopes can be
made arbitrarily shallow through violent relaxation processes. We will study
this aspect in this section by investigating the role of the exact value of the
initial slope of the single power-law profiles on the profile of the merger
remnant. Furthermore, we will investigate whether broken power law profiles
could serve as a natural end-point of such a process by being particularly
resilient to further mergers.

Indeed, the magnitude of the transformation depends sensitively on both the
number of mergers as well as on initial density profiles of the merging
systems. Greater numbers of mergers lead to larger potential fluctuations,
whereas steeper cusps have deeper potentials, which are more resilient to
potential fluctuations of similar magnitude. This can be clearly seen in
Fig.~\ref{fig:spheres}, where we plot the density profiles of merger remnants
whose progenitors had power-law density profiles with slopes of $\gamma=1$
(left), 1.9 (middle), and for Hernquist density profiles (right). As in
previous plots, different colours show results for different merger
configurations.

For both power-law models the  behaviour is qualitatively similar to the
$\gamma=1.5$ case discussed above: multiple equal-mass mergers result in
shallower remnant profiles, with the largest changes seen for cases with the
highest number of mergers. The relative change in inner slope appears to
correlate with the initial value: for $\gamma=1.0$, $1.5$, and $1.9$, the
fractional change in slope is $80$, $50$ and $30\%$, respectively. This is
because violent relaxation only occurs when fluctuations in the gravitational
potential are of the same order as the self-binding potential of the system,
which is substantially larger for systems with steeper inner density cusps. On
the other hand, changes to the Hernquist models are comparatively smaller. Even
for the quintuple merger scenario, the relative changes to the inner slope are
only $\sim$10\%, comparable to the statistical noise in our density estimates.
The stability of the Hernquist model to violent merging is consistent with
previous studies on the impact of mergers on the equilibrium structure of dark
matter haloes \citep[e.g.][]{Boylan-Kolchin2004,Kazantzidis2006,El-Zant2008}. 
Note that in the case of Hernquist or NFW profiles, the central potential is 
considerably deeper than in the single power-laws profiles considered above. 
The deeper potential implies that mergers induce smaller relative variations, 
which explains the smaller changes in the final density profile.

The above results emphasise the importance of properly modelling the initial
collapse of dark matter haloes near the free-streaming scale: {\it for
identical merger histories, halo profiles depend sensitively on the profiles of
their progenitors}. Additionally, a value of $-1$ for the inner slope does not
stand out as a special case in this context, and the final halo can, in
principle, acquire a range of values depending on the precise details of its
accretion history and progenitor population.


\section{Discussion and Conclusions}

Attempts to explain the origin of the universal broken power-law density
profiles (e.g. Eq.~\ref{eq:NFW}) of collisionless dark matter haloes have
permeated the literature on cosmological structure formation ever since their
discovery by \cite{Navarro1996,navarro1997}. A complete physical picture of
their origin, however, has not yet emerged. In this article, we approach this
subject from two angles: first from simulations of structure formation in a
neutralino-DM cosmology, and, second, from a set of well-controlled equal-mass
merger simulations of power-law haloes.

In a neutralino-DM universe, hierarchical growth begins at about an earth mass,
which is the mass of the very first haloes to collapse and the mass of the
smallest haloes to form at any cosmic epoch. Therefore, this situation provides
a well-defined scale to be resolved by numerical simulations, and thus a unique
test for the self-similarity of DM halo density profiles.  In this paper we
focus on the formation and evolution of six $\sim0.01-0.02\,\Msun$ haloes using
the high-resolution zoom technique. We simulate these haloes from $z=399$ to
$z=30$, tracing their mass-growth over more than four orders of magnitude, and
study the evolution of their density profiles and assembly histories. We chose
these haloes to reside in a not particularly dense environment inside of a
large cosmological volume in order to focus on typical microhaloes rather than
the high-sigma peaks.

In agreement with previous studies
\citep{Diemand2005b,Ishiyama2010,Anderhalden2013,Ishiyama2014}, we found that
this first generation of $\chi$DM haloes have spherically-averaged density profiles
that are well described by single power-laws, $\rho\propto r^{-1.5}$,
inconsistent with the gently curving NFW-profiles that are ubiquitous at later
times and at much larger halo masses. We then observed that, as our haloes grow in
mass, all of them experienced a significant reduction in their inner slope
(even becoming shallower than $-1$ in some cases). This transformation occurs
rapidly, and appears to coincide with an early evolutionary phase in which major
mergers are particularly abundant. The final outcome is that all of them display
an inner, shallower, slope close to $-1$ as well as a steeper outer profile. Even at
the latest times we considered, they are, while being close, however still not 
particularly well fit by NFW-profiles arguably owing to the rapid assembly phase
in which they still find themselves. 

Motivated by this, we explored the impact of major mergers on halo density
profiles using a suite of non-cosmological simulations of idealised major
mergers. We considered equal-mass mergers with a range of multiplicities (from
binary mergers up to mergers involving five identical haloes) as well as a
range of initial density profiles. We showed that, in all configurations,
temporal variations in the gravitational potential induced by the mergers lead
to a significant modification of the initial density profile. This is a
somewhat unexpected outcome, as it has been argued several times
\citep{Kazantzidis2006,Boylan-Kolchin2004,El-Zant2008} that mergers do not
change density profiles \citep[but see][]{Laporte2015}, but it is consistent
with the recent analysis of \cite{Ogiya2016}.

More specifically, we find that major mergers transform the entire halo profile
in the following ways: 

\begin{enumerate} 
\item A higher number of merger progenitors drives larger fluctuations in the
gravitational potential, resulting in more substantial reduction of the inner
density slope.
\item There is no clear limiting inner slope, and the final profile
can be significantly shallower than $r^{-1}$.  
\item Shallower progenitor profiles are more vulnerable to the effects of mergers, due to 
their lower self-binding energies.
\item Mergers establish an outer envelope which leads to a broken power-law
profile with a distinct inner and outer slope. 
\item If the progenitor profile is a broken power-law, the end result is a profile
whose shape is largely insensitive to its merger history (which make our findings
compatible with previous published results).
\end{enumerate} 

The general picture that emerges from these simulations is very clear: haloes
are born as single power law profiles with a slope close to $-1.5$ which then
evolve through major mergers to more complicated profiles with a shallower
inner and a steeper outer profile. Once a broken power law profile is
established, it becomes resilent to further mergers.

It is interesting to note that fluctuating potentials have been suggested as
the origin of cusp-core transformations in dwarf galaxies
\citep[e.g.][]{Zolotov2012,Pontzen2012,Ogiya:2014,El-Zant2016}, and that major
mergers can significantly alter the central structure of elliptical galaxies
\citep[e.g.][]{Hilz2012,Hilz2013}. Although the physical origin of fluctuating
potentials and mergers is completely different, the same physical processes
appear to be acting on shaping galaxies as well as subsolar DM haloes.

It is also interesting to conjecture about the Milky-Way-size descendant haloes
of neutralino haloes based on the evidence presented in this paper. We showed
that the initial collapse establishes a simple profile, $\rho\propto r^{-1.5}$.
These power-law profiles are unstable to strong perturbations and can be made
shallower by subsequent rapid major mergers. There is no reason preventing
haloes to emerge from the initial period of major mergers significantly
shallower than $-1$, however, the particular shape of the DM mass spectrum
appears not to allow enough mergers for this to happen to a typical halo. As
haloes grow in mass and the frequency of major mergers decreases, a more
persistent, broken power-law density profile emerges. Subsequent mass growth
and major and minor mergers are not expected to modify this profile much
further.  Therefore, we speculate that, in a neutralino-DM scenario, even
galactic-size haloes might show some variance in their inner density slope
--some of them being shallower than $-1$, but some of them steeper -- resulting
from the diversity of their early merger history. In the future, larger
statistical samples in larger simulation volumes, as well as over more extended
periods of mass growth, will be able to test the validity of this scenario.

We believe that the analysis we conducted in this article has important
implications for the interpretation of the universal structure of dark matter
haloes. However, several important questions remain open:

\begin{enumerate} 

\item How are the self-similar power-law profiles connected to the initial
(smooth) peak profile from which they form? Using different initial peak
profiles in numerical experiments it may be possible to connect the mass
profiles of these haloes to those of their initial (Lagrangian) peaks
\citep[see, e.g.][]{Vogelsberger2011}, and investigate the dependence of the
power-law index on the cut-off scale, power spectrum slope, and cosmology.

\item What is the dependence of the final halo on the cut-off scale? The
frequency and mass ratios of mergers that a typical halo will undergo during
its initial evolutionary phase are directly related to the slope and variance
of the power spectrum at the cut-off scale.  Simulations of the same objects
but embedded in different large-scale fluctuations will be able to investigate
in more detail the degree to which an initial profile can be transformed for a
given dark matter particle mass. 

\item Our results suggest that a simple over-density based criterion for
identifying haloes largely over-estimates their truly isotropised and
virialised regions at very early times. A robust definition of what constitutes
a virialised object, and how it can be identified and delineated reliably in a
simulation, should be explored in future studies of neutralino halo formation.

\item The most puzzling open question, however, is why standard CDM $N$-body
simulations so consistently produce haloes whose mass profiles resemble the NFW
profile? Future simulations with improved mass and force resolution, as well
as alternative methods with reduced noise, might be able to connect the apparently
contradictory results of neutralino-DM, WDM, and standard (perfectly cold) CDM simulations.
However, our results already indicate the importance of resolving the correct number of
major mergers in the early halo evolution.
\end{enumerate} 


We plan to address these questions in future work, and hope that our results
motivate others to investigate cosmological structure formation near the
free-streaming scale.

\section*{Acknowledgments}
\noindent
It is a pleasure to thank Tom Abel, Chervin Laporte, Thorsten Naab, Andrew
Pontzen, Justin Read, and Volker Springel for valuable discussions. REA
acknowledges support from AYA2015-66211-C2-2. This work was supported by a
grant from the Swiss National Supercomputing Centre (CSCS) under project ID
s616.

\bibliographystyle{mn2e} \bibliography{database}

\appendix
\section{Halo Centering and tracking}
In this Appendix, we discuss how we computed the centre of the cosmological
haloes from the N-body particles and tracked them between different snapshots
of our simulations.

Our resimulated haloes undergo very rapid mass growth and
experience multiple major mergers over a short timescale. This is problematic
when attempting to define halo centres, and more so when linking haloes to
their progenitors across simulation outputs. After some experimentation, we
found that algorithms commonly used in the analysis of CDM simulations failed
to yield satisfactory results under these extreme conditions. For instance,
during a major merger a shrinking sphere algorithm would sometimes converge to
the region between the two largest density peaks, and the position of the most
bound particle was sensitive to how mass was divided between the different
substructures, resulting in spurious temporal fluctuations in the recovered central
structure.

To circumvent these issues we define halo centres using a modified version of
the shrinking sphere algorithm of \cite{Power2003}. Our method uses an initial
guess for the location of the halo centre and identifies all particles with a
large radius $r_i$ centred on that point. We then smooth the density field of
these particles using a Gaussian kernel of size $0.1\times r_i$, and identify
its highest density maximum. The location of this peak is used as the new
centre, and the radius of the sphere decreased by $20\%$.  This procedure is
repeated until fewer than $1000$ particles are found within $r_i$, at which
point we adopt a traditional shrinking sphere algorithm. Using this centre, we
then compute the virial mass, $\Mvir$, and radius, $\Rvir$ of the halo.  

After computing halo centres in this way for a given simulation output,
we then identify the most likely descendant of each in the subsequent snapshot.
This is defined as the nearest halo in a hyperspace of position, virial mass,
$\Mvir$, and peak circular velocity, ${\rm V_{max}}$. More specifically, the 
distance between a halo $i$ at time $t_i$ and a possible
descendant halo $j$ in a snapshot at time $t_j$ is defined 

\begin{equation}
  d_{ij}^2 = |x(t_j)^i-x^j|^{2/3} + \log(\Mvir^i/\Mvir^j)^{2/3} + \log(v_{\rm max}^i/v_{\rm max}^j)^{2/3}, 
\end{equation}

\noindent where $x(t_j)^i = x(t_i)^i + (t_i-t_j) v^i(t_i) H(t_i)$ 
is the predicted position of the halo $i$ at time $t_j$. We found that this method
performed more robustly than simply tracking a fraction of the most bound
particles between simulation outputs, and gave consistent results during
quiescent phases of halo growth.

\section{Isolated halo particle sampling}
In this Appendix, we discuss how we set up the haloes for our isolated non-cosmological
merger simulations through which we study the evolution of halo density profiles during
and after a sequence of major mergers.

\subsection{Setting up halo profiles by distribution function sampling}
\label{sec:setting_up_profiles}
As discussed in Section~\ref{sec:dmhalomodels}, we consider spherically symmetric,
isotropic dark matter halo models with truncated power-law density profiles:

\begin{equation}
\label{eq:rho}
\rho(r) = \rho_s \left(\frac{r}{r_s}\right)^{-\gamma} \erfc[\kappa (r/r_s - 1)],
\end{equation}

\noindent For any spherically-symmetric mass distribution, the gravitational potential is 
given by \citep[][Eq 2.28]{Binney2008}

\begin{equation}
\Phi(r) = -\frac{G M(r)}{r} - 4 \pi G \int_r^{\infty} {\dd}r' r' \rho(r'),
\end{equation}
 
\noindent where $M(r)$ is the total mass enclosed within a radius $r$, and $G$
is Newton's gravitational constant.

The only ergodic distribution function that can lead to a spherically symmetric
system is given by Eddington's formula \citep{Binney2008}:

\begin{equation}
\label{eq:df}
f(E) = \frac{1}{\sqrt{8} \pi^2} \left[ \int_0^E \frac{{\dd}^2\rho}{{\dd}\Psi^2}  
       \frac{{\dd}\Psi}{\sqrt{E-\Psi}} + \frac{1}{\sqrt{E}} \left(\frac{{\dd}\rho}{{\dd}\Psi}\right)_{\Psi=0} \right],
\end{equation}

\noindent where $E \equiv -\Phi(r)-\frac{1}{2}v^2$ is the relative energy, $v$ is the
velocity, and $\Psi \equiv -\Phi$ is the relative gravitational potential. Note that since 
$\rho \rightarrow 0$ when $\Psi \rightarrow 0$, the second term in the above 
expression vanishes and Eq.~\ref{eq:df} can be rewritten as a function of only 
derivatives with respect to the radius:

\begin{eqnarray}
f(E) &=&  \frac{1}{\sqrt{8} \pi^2} \int_\infty^{r'(E)} \frac{{\dd}r}{\sqrt{E-\psi}} \\
&\times& \left[ \left( \frac{{\dd}\Psi}{{\dd}r}\right)^{-1} \frac{{\dd}^2\rho}{{\dd}r^2}-\frac{{\dd}\rho}{{\dd}r}\left(\frac{d\Psi}{{\dd}r}\right)^{-2}\frac{{\dd}^2\Psi}{{\dd}r^2}\right] ,
\end{eqnarray}

\noindent where
\begin{eqnarray}
\frac{d\Psi}{{\dd}r}     &=& G \frac{M(r)}{r^2} \\
\frac{d^2\Psi}{{\dd}r^2} &=& -2 G \frac{M(r)}{r^3} + 4 \pi G \rho(r).
\end{eqnarray}

\begin{figure}
\includegraphics[width=8cm]{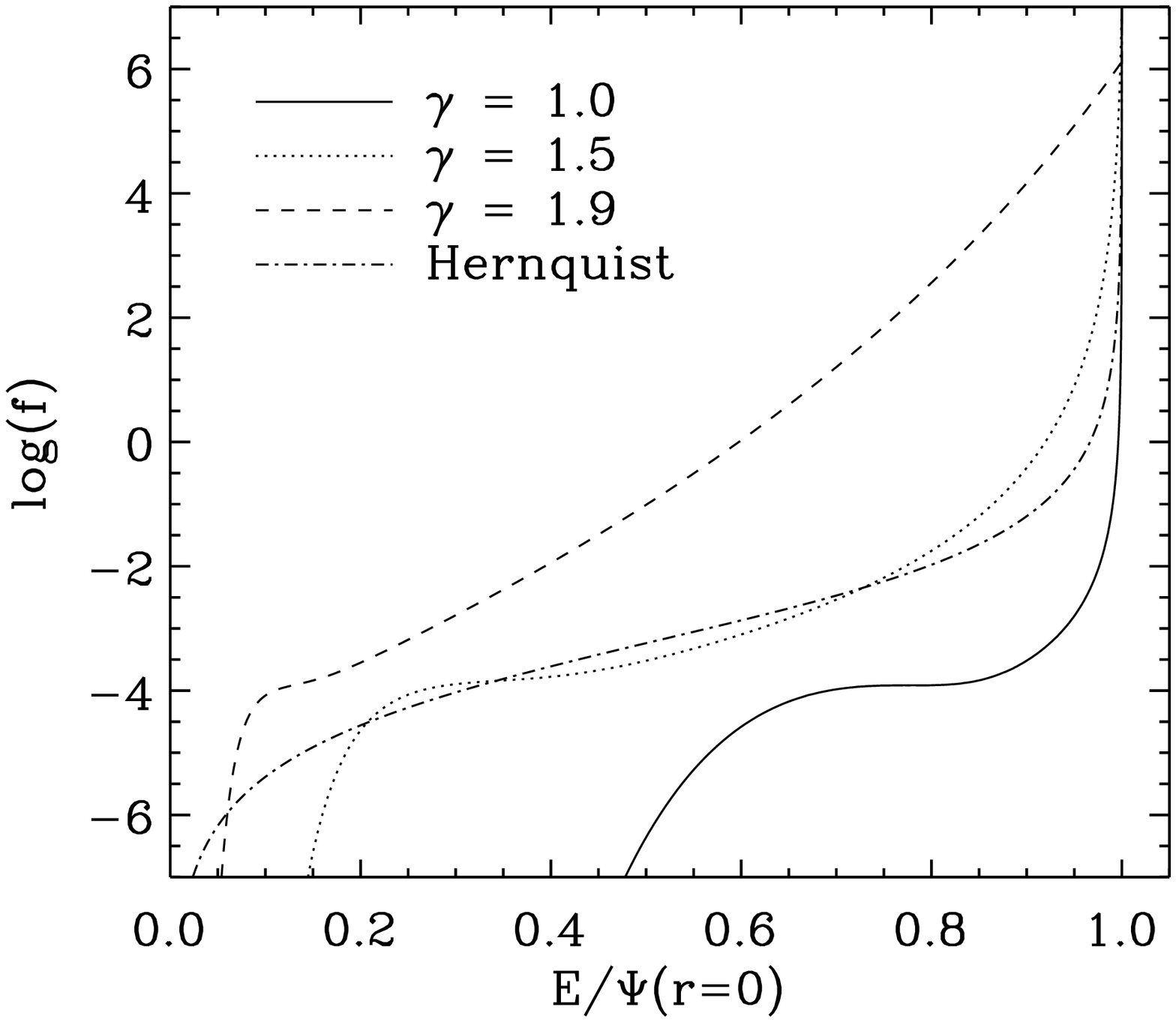}
\caption{Distribution functions for our halo models (Eq.~\ref{eq:rho}). 
Curves show results for three power-law indices, $\gamma = 1.0$
(solid line), $\gamma = 1.5$, (dotted line) $\gamma = 1.9$ (dashed line). 
The dot-dashed line show the distribution function for a Hernquist profile.}
\label{fig:df}
\end{figure}

\noindent The latter is more robust for numerical integration purposes. To solve for the
distribution function, we first compute and tabulate $M(r)$,
$\Psi(r)$, ${\dd}\rho/{\dd}r(r)$ and ${\dd}\rho^2/{\dd}r^2(r)$ using $2000$
logarithmically-spaced bins between $10^{-5}\,r_s$ and $10^3\,r_s$. We perform
the final integration using a Quadpack algorithm and interpolate the 
tabulated values when required. We note that an analytic solution for $f(E)$ is
known for the case of a Hernquist density profile \citep{Hernquist1990}. We use this
to test our numerical implementation, finding agreement to better than
$2\%$ over the whole range of allowed energies. We show the distribution
functions for the density profiles considered in this paper in Figure~\ref{fig:df}.

To create a stable initial configuration of particles, we first sample a
radius $r_i$ from the desired density distribution using an inverse transform
sampling. We then sample a velocity magnitude $v_i$ from the distribution 
$f(v){\dd}v = \sqrt{2 (\Psi(r_i) - f(E))} v{\dd}v $ using a Neumann rejection method.
The position and velocity of each particle are then assigned random orientations
in order to ensure the system's isotropy.

\subsection{Testing stability}
\label{sec:stability}

\begin{figure}
\includegraphics[width=8cm]{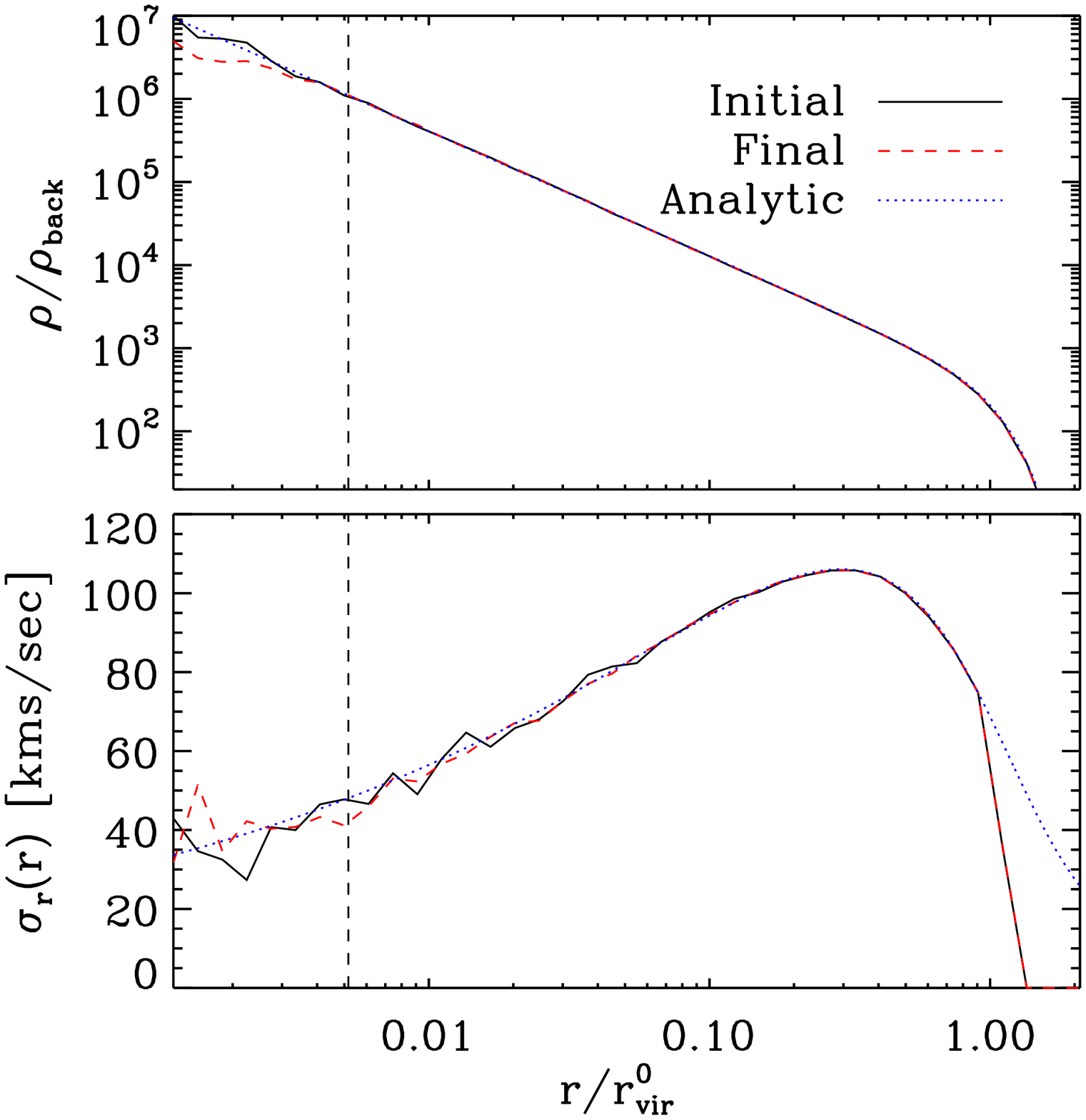}
\caption{Comparison between the initial (solid black lines) and final state
(dashed red lines) of an isolated dark matter halo model simulated with $10^6$
particles for $10$ Gyrs. The top panel show the measured density profiles
together with the analytic expression of Eq.~\ref{eq:rho}; the bottom panel
shows the radial velocity dispersion profile together with the solution of the
Jeans equation for a spherically symmetric and isotropic system (blue dashed lines). The
vertical dashed lines highlight $2.8\times\epsilon$, the Plummer equivalent
gravitational softening length used for our runs.\label{fig:single}}
\end{figure}

In order to assess the stability and accuracy of our models we created a single halo
with a ($\gamma=1.5$) power-law density distribution and evolved it in isolation. In Fig.~\ref{fig:single} we
show the initial and final density and velocity dispersion profiles. For
comparison, blue dotted lines show the analytic density profile and the
velocity dispersion profile computed using the Jeans equation for a spherical
system with an isotropic velocity distribution. We note that the simulated halo
agrees extremely well with the analytic results, showing small differences only in
the central regions where particle noise is not negligible. In addition, the
system is dynamically extremely stable: the final state agrees remarkably
well with the analytic results as well as with the initial configuration, with very
little departure over the entire range plotted. Small differences are 
most clearly seen in the very inner parts of the halo ($r < 0.005 r_{vir}$), where
the relaxation times are short and two-body scattering becomes important. However, this 
occurs on scales smaller than the formal resolution scale of our simulations. Therefore, 
these results validate our implementation as well as the integration and force calculation 
accuracy of our simulations.

\label{lastpage} 
\end{document}